**Evaluation des impacts cumulés sur la ressource marine d'un système socio-écologique corallien à Moorea (Polynésie Française) : approche par modélisation multi-agents**

# MÉMOIRE

**Présenté par :** Olivier Rousselle

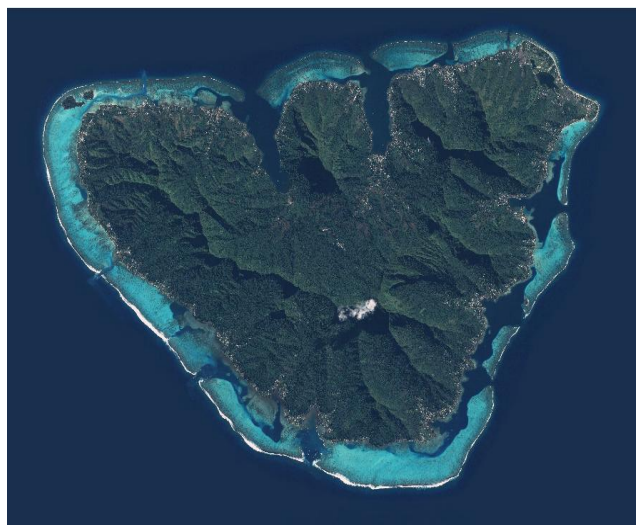

Pour l'obtention du :
Diplôme d'ingénieur d'AgroParisTech

Stage effectué du 01 / 05 / 2017 au 31 / 10 / 2017
A : CRIOBE, USR 3278 CNRS-EPHE-UPVD
    58 avenue Paul Alduy, 66860 Perpignan

Enseignant-responsable : Bruno Lemaire (AgroParisTech)
Maîtres de stage : Joachim Claudet (chercheur CNRS, CRIOBE)
                   Mélodie Dubois (doctorante CNRS, CRIOBE)
                   Patrick Taillandier (chercheur INRA)

Soutenu le : 14 septembre 2017

# Abstract

In a context of climate change and significant changes in human activities around the world, coral reefs are subject to many disruptions. In tropical island ecosystems, human activities (fishing and tourism in particular) are highly dependent on coral reefs and associated species. In Moorea, CRIOBE studies these interactions and actively participates in the management of these ecosystems (it contributed to the implementation of a Maritime Space Management Plan, PGEM). This PGEM is currently under revision, and consultations are underway between authorities and users of the lagoon space. In this context, we have developed a tool to help decision-making, it's an agent-based model. Different management scenarios have been tested (establishment of a quota, compliance with marine protected area regulations, night fishing ban, financial assistance to fishing capacity), and we have seen the repercussions on social (biomass fished, conflicts), and ecological (fish biomass in water, coral recovery) outputs. Each scenario was tested with or without environmental disturbance to assess the recovery capacity of coral ecosystems and the sensitivity of fish communities to scenarios.

We have modeled the trophic interactions with a Lotka-Volterra model, and the interactions between fishermen, trophic groups and tourist operators, over a period of 20 years, with alternation day-night. In order to parameterize the model, we used the CRIOBE ecological data, the ISPF social data, the results of scientific publications, and we carried out surveys among the population. The results were generated through global, temporal (time series), and spatial (GIS maps) outputs. Each scenario tested has advantages and disadvantages: the quota reduces overall fishing pressure, but brings more fishing pressure to the lagoon than to the fore reef; the theoretical absence of poaching ensures good regeneration of fishery resources in marine protected areas but is responsible for an imbalance in fishing pressure between these areas and others; the prohibition of night fishing allows a better state of the fishery resource, but increases the day conflicts; finally, financial aid ensures a better distribution of fishing pressure, with less pressure in the lagoon, but is more sensitive than other scenarios in case of an environmental disturbance (cyclone or starfish *Acanthaster plancii*).

The model produced here can be transposed to other ecological and economic situations, by modifying the parameters, and to other geographical areas, by changing the input map data.

**Key words** : coral reefs - interaction fishing-tourism - decision support tool - management scenarios - agent-based model.

# Résumé

Dans un contexte de changement climatique et de modifications importantes des activités humaines à travers le monde, les récifs coralliens sont sujets à de nombreuses perturbations. Dans les écosystèmes insulaires tropicaux, les activités humaines en mer (pêche et tourisme notamment) sont très dépendantes des récifs coralliens et des espèces associées. A Moorea, le CRIOBE étudie ces interactions, et participe activement à la gestion de ces écosystèmes (il a contribué à la mise en place d'un Plan de Gestion de l'Espace Maritime, PGEM). Ce PGEM est actuellement en cours de révision, et des concertations sont en cours entre autorités, et usagers de l'espace lagonaire. Dans ce cadre, nous avons développé un outil d'aide à la prise de décision, il s'agit du modèle multi-agents. Différents scénarios de gestion ont été testés (mise en place d'un quota, respect de la réglementation sur les aires marines protégées, interdiction de pêche la nuit, aide financière à la capacité de pêche), et nous avons vu les répercussions sur des sorties sociales (biomasse pêchée, conflits en mer), et écologiques (biomasse de poissons dans l'eau, recouvrement corallien). Chaque scénario a été testé avec ou sans perturbation environnementale afin d'évaluer la capacité de récupération des écosystèmes coralliens et la sensibilité des communautés de poissons en fonction des scénarios.

Nous avons modélisé les interactions trophiques à travers un modèle de Lotka-Volterra, et les interactions entre pêcheurs, groupes trophiques et opérateurs touristiques, sur une durée de 20 ans, avec alternance jour-nuit. Afin de paramétrer le modèle, nous avons utilisé les données écologiques du CRIOBE, les données sociales de l'ISPF, les résultats de publications scientifiques, et nous avons réalisé des enquêtes auprès de la population. Les résultats ont été produits à travers des sorties globales, temporelles (séries temporelles), et spatiales (cartes SIG). Chaque scénario testé présente des avantages et des inconvénients : le quota permet de diminuer la pression de pêche globalement, mais amène une pression de pêche plus importante dans le lagon que dans la pente externe ; l'absence théorique de braconnage assure une bonne régénération de la ressource halieutique dans les aires marines protégées, mais est responsable d'un déséquilibre de la pression de pêche entre ces zones et les autres ; l'interdiction de pêche de nuit permet un meilleur état de la ressource halieutique, mais augmente les conflits de jour ; enfin, l'aide financière assure une meilleure répartition de la pression de pêche, avec moins de pression dans le lagon, mais est plus sensible que les autres scénarios en cas de perturbation environnementale (cyclone ou étoile de mer *Acanthaster plancii*).

Le modèle produit ici est transposable à d'autres situations écologiques et économiques, en modifiant les paramètres, et à d'autres zones géographiques, en changeant les données cartographiques d'entrée.

**Mots clés** : récifs coralliens - interaction pêche-tourisme - outil d'aide à la décision - scénarios de gestion - modèle multi-agents

# Sommaire

# Remerciements

Joachim Claudet, maitre de stage, chercheur CNRS - CRIOBE

Mélodie Dubois, co-encadrante du stage, doctorante EPHE - CRIOBE

Patrick Taillandier, co-encadrant de stage, chercheur INRA Toulouse, département MIAT

Bruno Lemaire, tuteur de stage, chercheur AgroParisTech

Catherine O'Quigley, directrice des études et de la pédagogie – centre de Paris – AgroParisTech

Lauric Thiault, doctorant UPMC - CRIOBE

Peter Estève, technicien informatique EPHE - CRIOBE

Christophe Blazy, stagiaire CEFREM

# I. Introduction

Les récifs coralliens font partie des écosystèmes les plus complexes et les plus biologiquement diversifiés de la planète, et ils rendent de nombreux services écosystémiques aux populations insulaires : ils permettent une protection de la côte en limitant l'érosion, assurent aux populations une sécurité alimentaire, soutiennent un développement économique local à travers la pêche et le tourisme, et sont la base de nombreuses pratiques culturelles contribuant au bien-être des populations locales (Pascal et al., 2015 ; Cinner, 2014). Cependant, les récifs coralliens, à travers le monde, sont sujets à de nombreuses perturbations environnementales telles que des invasions d'étoile de mer corallivores *(Acanthaster plancii)*, des cyclones et des évènements de blanchissement. Ces perturbations brutales et globalement imprévisibles provoquent une diminution drastique du recouvrement corallien, qui se répercute ensuite sur l'ensemble de l'écosystème (Lamy et al., 2015*)*. Jusqu'à récemment, les récifs coralliens ont démontré leur capacité à faire face à ces perturbations en retournant à un état de recouvrement optimum de la surface (Jackson et al., 1992 ; Pandolfi et al., 2006 ; Salvat et al., 2008) ; cependant de nombreuses observations faites ces dernières décennies montrent que, combinées à des pressions anthropiques (tel que la surpêche ou la pollution), les récifs coralliens subissent des changements de régime irréversibles, en passant d'un état productif dominé par le corail vers un état indésirable dominé par les macro-algues (Hughes et al., 1994 ; Kittinger et al., 2015 ; Graham et al., 2013). Ils font partie actuellement des écosystèmes les plus menacés de la planète (Mumby et al., 2008).

La prise de conscience que les hommes font partie intégrante des écosystèmes a mené au développement d'une « gestion des pêches fondée sur les écosystèmes » (EBFM), promue par la FAO (ICES, 2005). Cette considération doit intégrer les interactions homme – environnement *(*Kittinger et al., 2012*)*, la résilience des écosystèmes (Graham et al., 2013), les activités humaines en mer (pêche et tourisme notamment), à l'aide données issues d'enquêtes locales (Leenhardt et al., 2015) et de relevés écologiques. Une bonne gestion se doit d'être multi-usage, concilier à la fois préoccupations sociales et durabilité des écosystèmes, et anticiper les éventuelles perturbations écologiques (*Acanthaster plancii* ou cyclone) ou économiques. Des exemples récents ont montré que des désastres environnementaux et économiques ont pu être minimisés, voir évités, grâce à des choix de gestion adaptés pris aux bons moments, avec l'implication des autorités et de la population (Pereira et al., 2009).

Comment réagissent les écosystèmes coralliens face aux pressions anthropiques, et quels scénarios de gestion peuvent être mis en place afin d'assurer une durabilité à la fois écologique et sociale de ce système ?

Comme cas d'étude, nous avons considéré l'île de Moorea, pour plusieurs raisons: le récif corallien de Moorea a une forte capacité de résilience, malgré les perturbations dues aux cyclones et à *Acanthaster plancii* (Adjeroud et al., 2009 ; Adam et al., 2014*)* ; l'activité de pêche fait partie intégrante de la culture locale ; les changements socio-économiques y sont importants : développement du tourisme et de l'économie locale (Leenhardt et al., 2015) ;  et la Polynésie française est considérée comme un des pays où la consommation moyenne par habitant en poissons est la plus importante du monde. La nécessité d'avoir une pêche durable et de perdurer cette tradition a conduit à la mise en place du Plan de Gestion de l'Espace Maritime (PGEM) avec des Aires Marines Protégées (AMPs). Ce dernier a soulevé une opposition de la part de la population, et son efficacité écologique est en deçà des attentes initiales (Thiault et al., 2016). Depuis plus d'un an, une concertation est en cours avec les pêcheurs et les opérateurs touristiques pour réviser ce plan de gestion, et notre étude vise à produire un outil d'aide à la décision dans le cadre de cette révision.

Afin de répondre à ce problème, nous avons développé un modèle multi-agents pour modéliser la dynamique écologique et sociale de l'île de Moorea. Les modèles multi-agents sont de bons outils pour réaliser des prédictions et tester des scénarios de gestion ; cependant, les modèles écosystémiques qui intègrent les dynamiques écologiques, les perturbations environnementales liées au changement climatique, les activités anthropiques et des scénarios de gestion sont sous-développés actuellement (Weijerman et al., 2016, Johnson et al., 2013). Dans notre modèle, nous avons simulé l'évolution des ressources, de la biomasse pêchée et des conflits entre pêcheurs et activités touristiques, avec et sans perturbation environnementale, sur une durée de 20 ans. Nous avons comparé l'évolution de ces indicateurs autour de l'île entre le statu quo (référence) et 4 scénarios envisagés dans le cadre de la révision du PGEM : meilleure surveillance (i.e. absence de braconnage dans les aires marines protégées), mise en place d'un quota de pêche, interdiction de pêche nocturne, et aide financière à la mobilité.

# II. Matériel et méthodes

## II 1. Cas d'étude de Moorea

### II. 1. a) Situation géographique et particularités géomorphologiques

L'île de Moorea est située en Polynésie française, plus précisément dans les îles du Vent (figure 1). Elle possède une superficie de 133 km². Comme toutes les îles et atolls de Polynésie, Moorea est une île volcanique. Elle s'est formée à partir d'un point chaud qui s'est déplacé au cours du temps relativement au plancher océanique. L'île de Moorea possède du relief, avec un point culminant à 1207 m. (mont Tohiea). Le climat de Polynésie est de type tropical humide, avec une saison chaude et humide (de novembre à avril), et une saison fraîche et moins pluvieuse (de mai à octobre).

Au cours du temps, le récif corallien se développe, tandis que le plancher océanique s'affaisse. Lorsque le récif corallien est accolé à la côte, on parle de récif frangeant ; lorsqu'il est séparé de la côte par un espace, on parle de récif barrière, et cet espace s'appelle le lagon. Moorea est actuellement en phase de récif barrière, et évolue vers le stade d'atoll (où l'île se retrouve totalement immergée, ne laissant plus émerger que la couronne récifale). Andréfouët et al., 2005

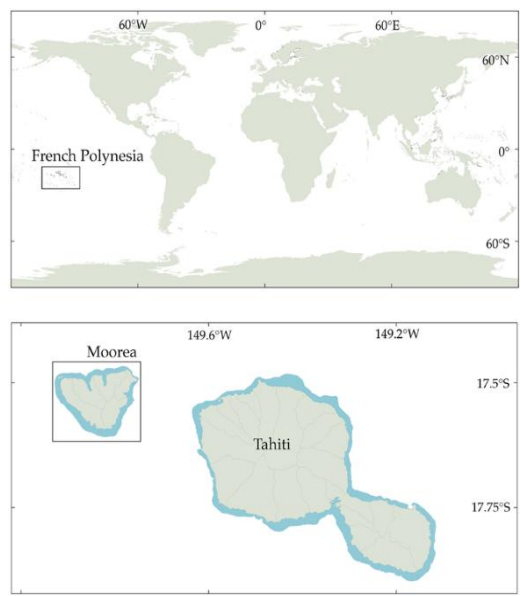

Figure 1 : Localisation de Moorea

Le lagon de Moorea a une surface de 49 km², sa largeur varie de 500 à 1500 m. et sa profondeur varie de 50 cm à 30 m. (Bell et al., 1984).

Au-delà de la barrière de corail, on trouve une forte pente qui fait la transition entre le lagon récif corallien et la zone pélagique océanique : c'est la pente externe (Andréfouët et al., 2005). Le lagon et la pente externe qui sont deux habitats distincts, aux fonctionnalités différentes, ont été considérés dans cette étude (figure 2). La récif barrière protège la zone lagonaire de l'océan, nous voyons sur la figure 2 les vagues qui viennent « taper » contre le récif barrière.

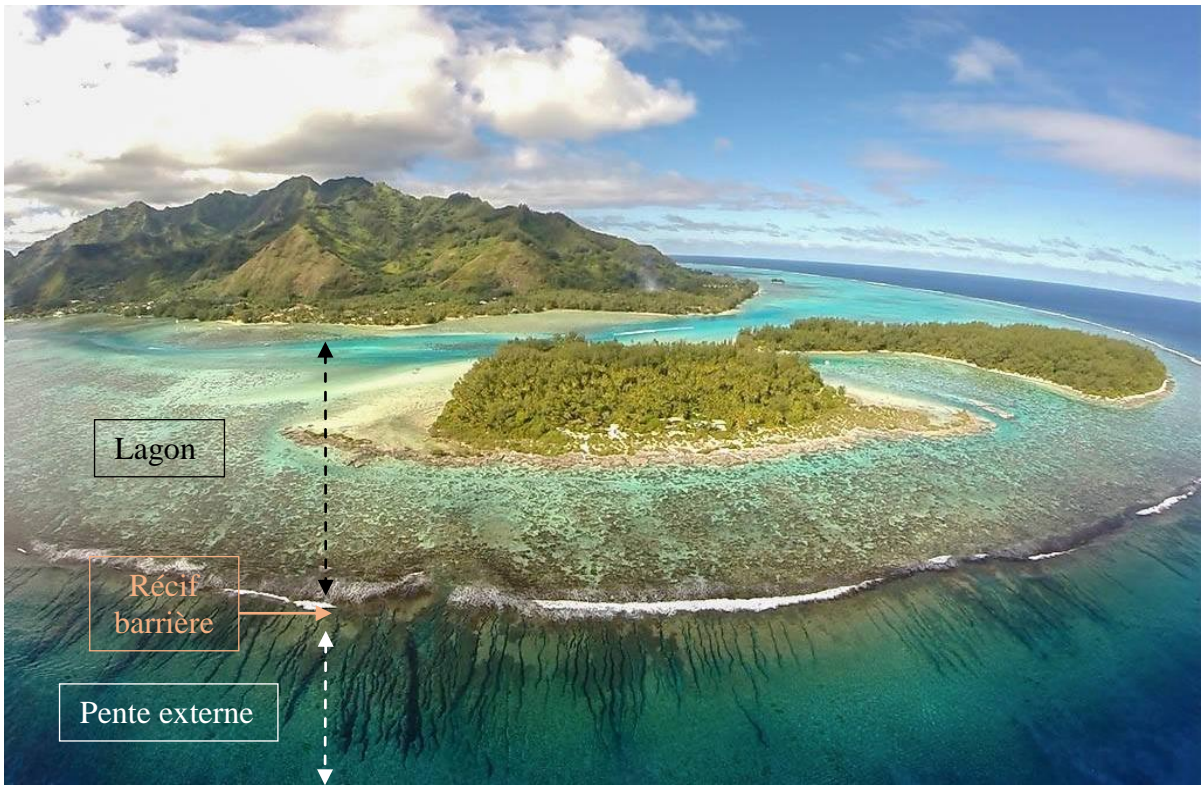

Figure 2 : Vue aérienne de la zone de Tiahura, Moorea, permettant de distinguer le lagon de la pente externe. Source : Antoine Pouget, CRIOBE

## II. 1. b) Fonctionnement écosystémique et perturbations environnementales

Moorea possède une très riche biodiversité marine, avec plus de 500 espèces de poissons. Le substrat, quant à lui, est composé de coraux (massifs et branchus), d'algues (turfs et macro-algues notamment), et de sédiments. Le turf correspond à l'ensemble des petites pousses formant du gazon algale, tandis que les macro-algues sont des algues de grande taille. Les algues et les coraux entretiennent une relation de compétition, afin d'occuper la surface. Parmi les poissons, on distingue les groupes herbivores (qui mangent des algues), les corallivores (mangent les coraux), et les carnivores (qui mangent d'autres poissons). De nombreuses espèces de poissons dépendent des récifs coralliens, comme source de nourriture ou comme refuge (Wilson et al., 2006). A Moorea, l'état des poissons de récif se portent relativement bien actuellement, aucune diminution importante n'a été observée sur la dernière décennie (M. Dubois, données CRIOBE ; Salvat et al., 2008) ; ceci est à mettre en perspective au niveau mondial, où on estime qu'un quart des ressources sont surexploitées, épuisées ou en phase de récupération (FAO, 2007).

Tout comme les autres îles de Polynésie, Moorea est touchée de façon cyclique, par deux types de perturbation environnementale (Lamy et al., 2015, Salvat et al., 2008) :

- Les phénomènes cycloniques, qui n'affectent également que les coraux de la pente externe, les autres étant protégés par le récif barrière. Ils provoquent une destruction de la structure du corail. Le dernier cyclone date de 2010, il s'agit du cyclone Oli.

- Des évènements d'invasion d'étoile de mer corallivore, *Acanthaster plancii*. Venant du domaine océanique, cette dernière se nourrit des polypes (partie vivante organique) des coraux en pente externe, entraînant une importante et rapide diminution du recouvrement corallien vivant (figure 3). Ces étoiles de mer meurent quand il n'y a plus de corail à manger. On la désigne également par l'abréviation "COTS" (pour Crown-of-thorns seastar en anglais). Cette régression du corail s'accompagne d'un développement des algues (Ainsworth et Mumby, 2015). La dernière invasion d'*Acanthaster plancii* date de 2006.

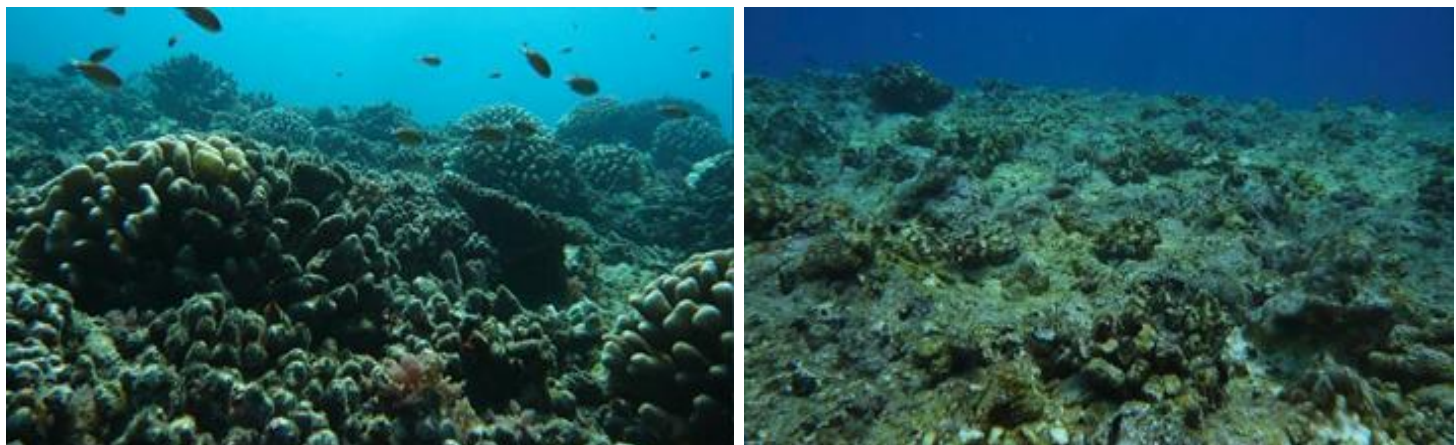

Figure 3 : Illustration de l'évolution de la couverture corallienne sur la pente externe de Moorea avant (à gauche) et après (à droite) passage d'*Acanthaster planci* et d'un cyclone.
Zone de Vaipahu. Source : CRIOBE

Ces perturbations ne touchent pas seulement les récifs coralliens, mais influencent aussi de façon indirecte les communautés de poissons de récif (Wilson et al., 2006). Malgré toutes ces perturbations, le récif corallien de Moorea s'est montré particulièrement résilient jusqu'à présent en retournant à son état initial en une dizaine d'années (figure 4), ce qui représente un recouvrement optimum 50–60 % de la surface (Salvat et al., 2008). Ces perturbations naturelles sont les perturbations majeures à Moorea, et leur fréquence d'apparition va probablement augmenter avec le réchauffement climatique ayant potentiellement des impacts sur les coraux et les communautés de poissons associés (Adam et al., 2014).

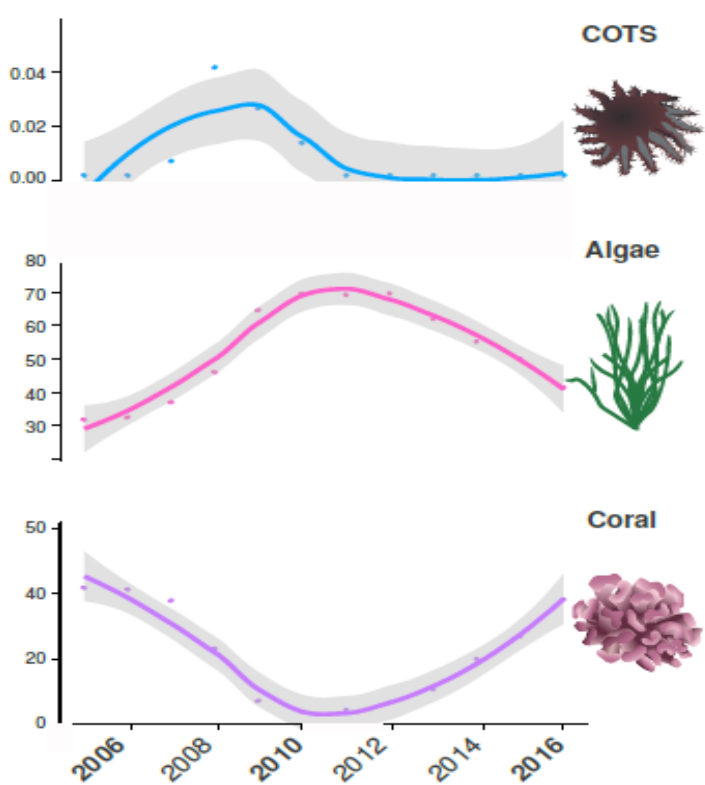

- 2006 : arrivée d'*Acanthaster plancii*
- 2006 – 2011 : invasion des étoiles de mer et mort des coraux de la pente externe
- 2010 : cyclone Oli
- 2011 : corail à son taux de recouvrement minimal
- 2011 – 2016 : récupération du corail
- 2016 : corail revient à son taux de recouvrement optimum

Figure 4 : Evolution du pourcentage de recouvrement en corail vivant, en macro-algues et Acanthaster sur la pente externe de Moorea.

Source : M. Dubois, données CRIOBE

### II. 1. c) Contexte socio-économique

Moorea a une population de plus de 17 000 habitants. L'économie de l'île, autrefois principalement basée sur l'agriculture et la pêcherie familiale, est maintenant basée sur l'agriculture d'exportation, le tourisme (avec des hôtels de luxe comme des pensions de famille), et les activités marines (plongée, balades en mer, nourrissage de requins). Moorea est l'une des îles les plus visitées de Polynésie française, avec plus de 70 000 touristes annuellement, ce qui occasionne des impacts importants sur les écosystèmes coralliens. (Leenhardt et al., 2016).

La pêche, autrefois moyen de subsistance essentiel avec l'agriculture, est maintenant plus une pratique de loisir, tout en gardant sa valeur culturelle. En effet, avec le développement économique de l'île, les ménages sont moins dépendants de la pêche pour survivre. Parmi les pêcheurs de Moorea, 20% ont pour motivation de pêche la subsistance ; 11 % pratiquent la pêche pour le commerce ; et 69 % pour le loisir. Les pêcheurs représentent environ 25% de la population globale de Moorea. Les activités de pêche à Moorea sont très dispersées, à la fois spatialement autour de l'île (on distingue la pêche récifale, la pêche côtière et la pêche dans l'océan) et temporellement (avec des pêcheurs de jour et des pêcheurs de nuit). On peut qualifier la pêche à Moorea « d'artisanale » ou « petite pêcherie côtière », elle a une importante culturelle très importante et contribue à l'identité de la population locale (au même titre par exemple que la langue tahitienne). Cette petite pêcherie côtière a des effets bénéfiques pour les communautés locales, mais peut impacter les ressources sur le long-terme si elle n'est pas gérée correctement. (Leenhardt et al., 2016).

### II. 1. d) Plan de gestion de l'espace maritime

L'île de Moorea a mis en place, depuis 2004, un Plan de Gestion de l'Espace Maritime (PGEM). Ce dernier, qui est en cours de révision depuis 2016, a quatre objectifs : l'utilisation rationnelle et la valorisation des ressources et de l'espace ; la gestion des conflits d'utilisation du lagon ; le contrôle des pollutions et des dégradations du milieu marin ; la protection des écosystèmes marins et des espèces menacées (Leenhardt et al., 2012). Malgré les processus de consultation de la population, ce plan est encore actuellement une source de tensions et de controverses (Walker et al., 2009).

Dans le cadre de ce plan, 8 aires marines protégées (AMPs) ont été mises en place afin de protéger et de pérenniser les ressources halieutiques ; la pêche y est interdite et elles couvrent environ 20% de la surface totale du lagon. A travers le monde, les AMPs ont été promues comment un élément central de la gestion durable, avec comme effets potentiels une augmentation des stocks de poissons, une augmentation du recrutement dans les zones autour des AMPs par spillover et une meilleure résilience des écosystèmes en cas de perturbation (Doyen et al., 2007).

Les données cartographiques / spatiales de l'île sont présentées dans la figure 5 ci-dessous. Les fichiers SIG sont des données issues du CRIOBE. Les différents compartiments de l'étude sont représentés : lagon / pente externe ; zones marines AMP / hors AMP ; ménages et districts à terre. Le lagon correspond à la partie marine (zone en bleu clair) entre le trait de côte et la pente externe. Les passes sont les zones où le récif barrière est « interrompu » et où le lagon communique avec l'océan. Les pêcheurs qui veulent se rendre en pente externe doivent passer par les passes, car hors de ces passes, le récif barrière leur empêche le passage.

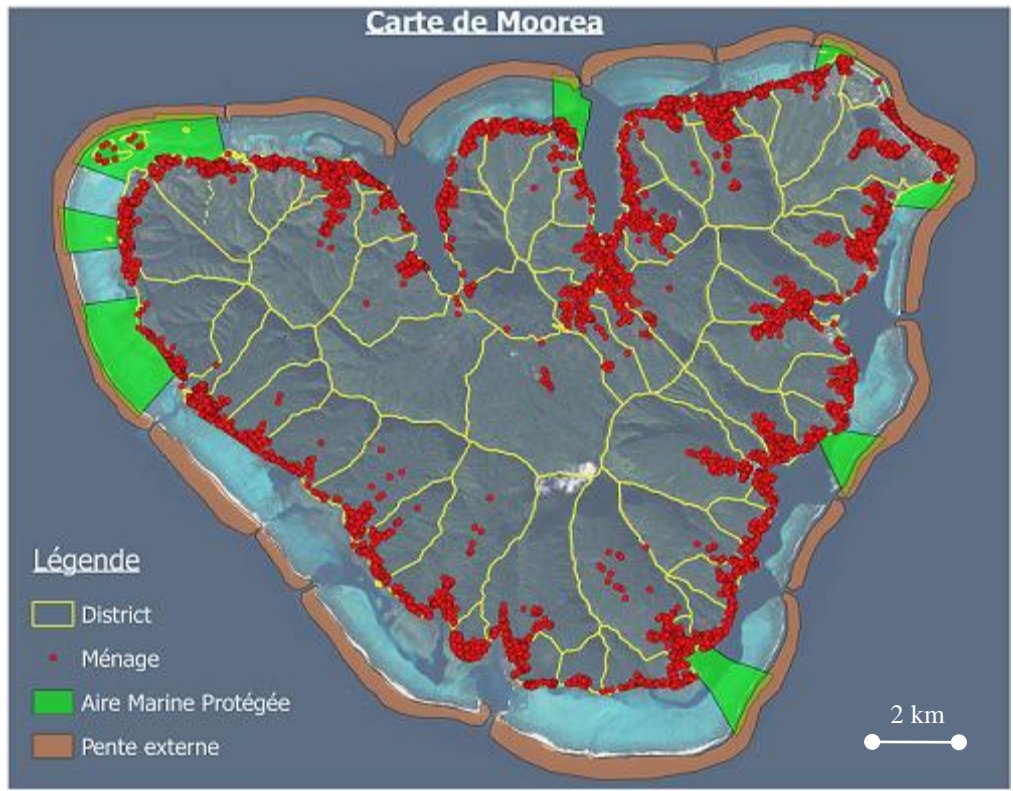

Figure 5 : Carte de Moorea représentant les différentes unités de l'étude

## II. 2. Présentation du modèle multi-agents

La programmation du modèle s'est faite à l'aide du langage multi-agents Gama. Gama a été développé par l'IRD (Institut de Recherche pour le Développement) afin de réaliser des simulations « spatialisées ».

La résolution utilisée est de 100m x 100m (soit 10 000 m² ou 1 ha). Cette taille correspond à une case unitaire dans notre modèle spatial. Le pas de temps est de 12 heures (une demi-journée), afin de prendre en compte les alternances jour-nuit. La simulation démarre en 2017, sur une durée de 20 ans.

Nous présentons sur la figure 6 le diagramme des interactions entre les différents agents du modèle (écologiques et sociaux), ainsi que la nature de leurs relations. Nous distinguons les agents trophiques (poissons herbivores, corallivores, carnivores), benthiques (corail et turf), et humains (pêcheurs et opérateurs touristiques).

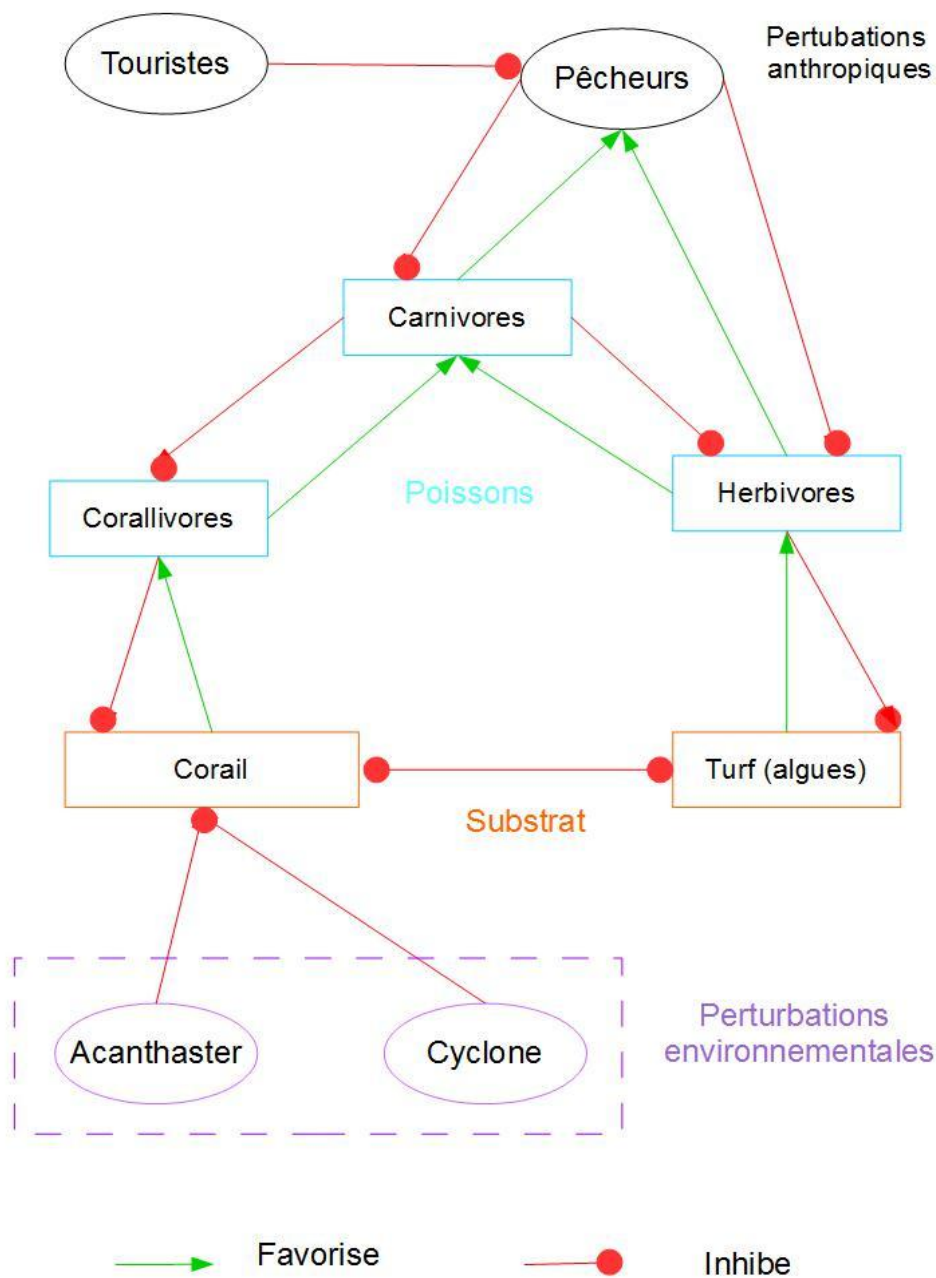

Figure 6 : Présentation des agents du modèle et de leurs interactions

## II. 2. a) Compartiment écologique

- <u>Groupes fonctionnels – agents :</u>

Le corail structure le récif, et les algues sont en compétition avec le corail. Parmi ces algues, nous avons considéré le turf et les macroalgues. Les algues et les coraux sont les espèces benthiques du modèle, leur abondance est exprimée en pourcentage de recouvrement de la surface du substrat. Nous considérons ici que l'abondance des macroalgues ne varie pas au cours du temps, à la différence du turf qui est fortement influencé par la présence de corail et d'herbivores (Adjeroud et al., 2002).

Les poissons jouent des fonctions clés, d'herbivorie, corallivorie et carnivorie. Ces groupes trophiques constituent les espèces pélagiques du modèle, leur abondance est exprimée en biomasse (kilogrammes).

- <u>Interactions écologiques : modèle de Lotka-Volterra</u> :

Sources : Y. Takeuchi, 1996 ; Tribut et al., 2013

La dynamique écologique (poissons, turfs et coraux) est régie ici par des interactions de type Lotka-Volterra, généralisées à plus de deux agents.

Dans nos équations de Lotka-Volterra généralisée, chaque agent présente une équation de ce type :

$$\frac{dx}{dt} = x * (\alpha_x - \gamma_x - \beta_{xy} * y + \delta_{xz} * z) \quad (1)$$

$x$ : biomasse de l'agent (en kg) ; $\frac{dx}{dt}$ : taux de variation de $x$ (en kg / j.)

$\alpha_x$ : taux de croissance intrinsèque (en jour$^{-1}$) ; $\gamma_x$ : taux de mortalité naturelle (en j$^{-1}$)

$y$ : biomasse de (ou des) espèce(s) inhibitrices (rôle de prédateur ou de compétiteur) ; en kg

$\beta_{xy}$ : taux de mortalité / inhibition de l'agent $x$ due à $y$ (en jour$^{-1}$. kg$^{-1}$)

$z$ : biomasse de (ou des) espèces(s) favorisantes (rôle de proie ou relation de commensalisme ou mutualisme). En kg.

$\delta_{xz}$ : taux de croissance due à l'agent $z$ (en jour$^{-1}$. kg$^{-1}$).

Les unités de biomasse (kg) prises ici s'appliquent au cas des poissons. Pour les coraux et les algues, les équations sont les mêmes, sauf que l'on parle ici en pourcentage de recouvrement de la surface.

Dans le tableau 1, nous présentons les interactions entre les différents agents écologiques qui sont inclus dans la dynamique de Lotka-Volterra. En colonne (en bleu) sont indiqués les agents, et en ligne (en gris) les agents qui les influencent.
- Un effet favorisant est indiqué par un signe $+$ ; nous avons alors le terme ( $+ \delta * z$ ) dans l'équation (1) de l'agent x.
- Un effet inhibant est indiqué par un signe $-$ ; nous avons alors le terme ( $- \beta * y$ ) dans l'équation (1) de l'agent x.
- Un effet nul / neutre est indiqué par un signe 0 ; l'agent n'apparait pas dans l'équation (1).

Tableau 1 : nature des interactions écologiques entre les agents du modèle
Sources : Lamy et al., 2015 - Leenhardt et al., 2017 – données CRIOBE

| Impactant \ Impacté | Corail | Turf (algues) | Herbivores | Corallivores | Carnivores |
|---|---|---|---|---|---|
| Corail |  | − | 0 | + | 0 |
| Turf (algues) | − |  | + | 0 | 0 |
| Herbivores | 0 | − |  | 0 | + |
| Corallivores | − | 0 | 0 |  | + |
| Carnivores | 0 | 0 | − | − |  |
| Pêcheur | 0 | 0 | − | 0 | − |
| *Acanthaster plancii* | - | 0 | 0 | 0 | 0 |

Les coraux et le turf ont un effet mutuel négatif : c'est une relation de compétition pour occuper la surface. Quand il y a moins de coraux, les algues se développent, et inversement.

A noter que les pêcheurs et *Acanthaster plancii* n'ont pas d'équation de Lotka-Volterra spécifique, mais ils influencent négativement des agents du modèle.

Si l'agent a uniquement le rôle de proie (corail et turf ici), on n'a pas les termes γ (taux de mortalité intrinsèque) et ( $+ \delta * z$ ) dans son équation descriptive. On retombe alors sur l'équation classique de Lotka-Volterra pour une proie.

Le corail est influencé négativement par le turf, les corallivores et *Acanthaster plancii*.
Son équation descriptive est donc la suivante :

$$\frac{dC}{dt} = C * (\alpha_C - \beta_T * T - \beta_{CO} * CO - coeff_{destruction_{COTS}} * COTS)$$

$C$ : couverture en corail (%) ; $\alpha_C$ : taux de croissance intrinsèque du corail (j$^{-1}$) ;

$T$ : couverture en turf (%) ; $\beta_T$ : coefficient de mortalité à cause de la présence des turfs (j$^{-1}$)

$CO$ : biomasse de corallivores (kg) ; $\beta_{CO}$ : coefficient de mortalité à cause des corallivores (j$^{-1}$.kg$^{-1}$)

COTS : indicateur de présence d'Acanthaster : 0 si absence ; 1 si présence (perturbation écologique)

$coeff\_destruction\_COTS$ : coefficient de destruction du corail par Acanthaster.

L'ensemble des équations de Lotka-Volterra pour les différents agents sont indiquées en annexe 1.

Equilibre de Lotka-Volterra :

Nous avons considéré que le système (écosystèmes + pêche) est à l'équilibre. Afin d'assurer cet équilibre initialement, nous avons paramétré le modèle grâce aux hypothèses suivantes :

$$\begin{cases} \alpha = \gamma \text{ (taux de croissance intrinsèque égal au taux de mortalité intrinsèque)} \\ \beta = \dfrac{\alpha}{y_0} \quad ; \quad y_0 : y \text{ à l'état initial} \\ \delta = \dfrac{\gamma}{z_0} \quad ; \quad z_0 : z \text{ à l'état initial} \end{cases} \quad (2)$$

- Dynamique du substrat :

Le recouvrement maximal du corail et du turf a été fixé. Pour chaque case, nous avons les recouvrements en turf et en corail (ce sont les données initiales). Tout au long de la modélisation, nous avons, à chaque pas de temps t et par case k :

$$\text{Corail}_{k,t} + \text{Turf}_{k,t} \leq \text{Corail}_{k,t_{initial}} + \text{Turf}_{k,t_{initial}}$$

- Dynamique des poissons :

Pour les poissons, nous considérons une capacité de charge (ou carrying capacity, correspondant à la biomasse maximale qui peut être supportée par la capacité environnementale). Nous la définissons pour chaque groupe de poissons et par case. La capacité de charge pour chaque pixel et pour chaque groupe de poisson a été estimée comme :

$$carrying\_capacity = (Biomasse\ initiale) * 2$$

En effet, il a été observé que lorsqu'une aire marine protégée est mise en place sur une zone, la biomasse de cette zone se voit en moyenne doublée. (Lester et al., 2008)

De plus, nous avons introduit un coefficient de spillover pour rendre compte du déplacement des poissons entre les cases. Ce dernier a été fixé à 0,9 suggérant que lorsque la biomasse de poissons atteint 90% de sa capacité de charge, 10% (100% - 90%) de cette biomasse initiale est déplacée aux cases voisines. C'est le phénomène de spillover.

De plus, nous enlevons également la biomasse capturée par les pêcheurs (c.f. partie II. 2. b), paragraphe « captures »). Seuls les carnivores et les herbivores sont pêchés dans notre modèle.

- Perturbation environnementale :

On souhaite introduire une perturbation environnementale dans le modèle afin de représenter au mieux la réalité. Il est nécessaire d'intégrer les approches spatiales et temporelles, afin d'explorer les effets de ces perturbations.

Comme vu précédemment, l'étoile de mer *Acanthaster plancii* et les cyclones ont un impact négatif sur les coraux de la pente externe. Les étoiles de mer se déplacent par fronts, et il leur faut environ 5 ans pour détruire entièrement l'ensemble des coraux de la pente externe de Moorea. Les coraux, eux, mettent 5 ans pour récupérer en moyenne (figure 4).

Dans notre modèle, nous nous intéressons aux conséquences d'une perturbation environnementale sur le corail, sans distinguer s'il s'agit d'une invasion d'*Acanthaster plancii* ou d'un cyclone (on considère que ces deux perturbations ont le même effet). La simulation dure 20 ans, et la perturbation intervient à t = 2 ans afin de laisser suffisamment de temps au système pour récupérer.

## II. 2. b) Compartiment halieutique

- <u>Variables associées à la pêche</u>

Afin d'avoir un ordre de grandeur des valeurs des différents paramètres sociologiques, nous avons réalisé un questionnaire à destination des pêcheurs de Moorea. Ce questionnaire a été diffusé sur les groupe Facebook "Mooz select fishing" (regroupant des pêcheurs de Moorea), et est présenté en annexe 5. Nous l'appellerons « questionnaire Select fishing ». Les pêcheurs interviewés exercent la pêche en activité secondaire pour la plupart.

<u>Informations fournies à l'échelle du ménage</u> : A partir du questionnaire ISPF, à l'échelle de l'individu (annexe 3) ou du ménage (annexe 4), nous sommes capables de retrouver par ménage le nombre de pêcheurs professionnels, le nombre de pêcheurs annexes, le nombre de bateaux, le nombre de pirogues. Les pêcheurs annexes exercent la pêche en activité secondaire. En se basant sur les données ISPF, nous avons un total de 2244 pêcheurs, dont 440 pêcheurs pro et 1804 pêcheurs annexes.

On ne connait pas la localisation précise du ménage qui a répondu au questionnaire ISPF, mais nous savons à quel district il appartient et nous disposons d'une couche SIG des localisations des ménages. Ainsi notre ménage se localisera au hasard dans un ménage du district correspondant.

Chaque pêcheur est défini par 7 paramètres : rayon de pêche, durée de pêche, jours de pêche, sélectivité, période de pêche et probabilité de braconnage, taux de capture. Ces paramètres sont résumés dans le tableau 2.

- Distinction pêcheurs professionnels / pêcheurs annexes : les pêcheurs professionnels se voient attribuer une **durée de pêche** de 8h, tandis que les pêcheurs annexes se voient attribuer une durée de 4h.

  La distinction se porte également sur les jours de pêche. Dans le modèle, chaque pêcheur a une **probabilité d'aller pêcher**, à chaque journée. D'après les travaux de Leenhardt et al., 2016, les pêcheurs commerciaux réalisent 2 à 5 sorties de pêche par semaine, tandis que les pêcheurs annexes en réalisent 1 à 3. Les pêcheurs pro se voient donc attribuer une probabilité de pêche qui est un nombre au hasard entre $\frac{2}{7}$ et $\frac{5}{7}$ ; entre $\frac{1}{7}$ et $\frac{3}{7}$ pour les pêcheurs annexes.

- **Sélectivité** : Chaque pêcheur se voit attribuer une sélectivité au hasard (nombre entre 0 et 1). Un pêcheur sélectif pêchera des espèces en particulier, tandis qu'un pêcheur non sélectif ne fait pas de "tri". Nous considérons ici que plus un pêcheur est sélectif, moins il pêche en biomasse.

- **Période de pêche (jour ou nuit)** : Dans notre modèle, nous considérons une alternance jour-nuit. Un pêcheur se voit attribuer au hasard une période de pêche : soit le jour (probabilité de 30%), soit la nuit (probabilité de 70%). En effet, à Moorea, il y a plus de pêcheurs de nuit que de pêcheurs de jour.

- **Rayon de pêche** : le rayon de pêche traduit la mobilité du pêcheur. 3 cas possibles :

  o Le ménage a accès à un bateau -> les pêcheurs de ce ménage ont un rayon de pêche de 10 km.

  o Le ménage a accès à une pirogue mais pas de bateau -> le rayon des pêcheurs est de 3 km.

  o Le ménage n'a ni accès à une pirogue ni à un bateau (pêche à la nage) -> rayon de pêche de 1 km.

- **Dépendance à la ressource** (Thiault et al., 2017) : l'indice de dépendance à la ressource a été calculé à partir des données ISPF pour tous les foyers de l'île. Il est fonction du nombre de foyer ayant la pêche comme activité de subsistance, du nombre de foyers ayant une activité de subsistance non liée à la pêche et du nombre de foyers ayant des personnes inactives à charge. Ce paramètre est important car, selon les travaux de Cinner et al., 2012, la surexploitation est fortement influencée par la dépendance à la ressource. Une forte dépendance à la ressource est également synonyme d'une grande vulnérabilité des populations locales (Kronen et al., 2010). Ici, la dépendance à la ressource est une valeur par district, entre 0 et 1.

    **Probabilité de braconnage** : nous avons considéré que la propension d'un pêcheur à braconner dépend de la dépendance à la ressource du district, de sa période de pêche et de la surveillance de la part des autorités. Elle correspond à la probabilité qu'a un pêcheur d'aller pêcher en zone AMP (Aire Marine Protégée). Le paramètre de surveillance a été fixé à 20%, signifiant que le braconnage diminue d'un facteur 1,2 grâce à la surveillance. La nuit, nous avons considéré que les pêcheurs braconnent fortement, la probabilité de braconnage vaut 1 / (1 + surveillance). Le jour, la probabilité augmente avec la dépendance à la ressource du district et est inférieure à celle de nuit : dependance_ressource_district / (1 + surveillance).

- **Taux de capture** : les pêcheurs ont plus de facilité à capturer du poisson la nuit que le jour. C'est ainsi que nous avons introduit un coefficient de capture : Le taux de capture le jour a été choisi de telle sorte qu'un pêcheur pêche environ 5 kg par jour en moyenne. Cette valeur de 5 kg a été obtenu à partir des résultats du questionnaire « select fishing », bien que les travaux de Leenhardt et al., 2016 montrent que les rendements de pêche sont difficiles à estimer à Moorea. La valeur du taux de capture jour a été fixée à 0,002 $j^{-1}$. La nuit, il est deux fois plus important que le jour.

Tableau 2 : Paramètres relatifs à chaque pêcheur

| Paramètre | Unité / valeur | Source / obtention |
|---|---|---|
| Rayon de pêche | 1, 3 ou 10 (km) selon l'équipement | ISPF, à l'échelle du ménage Questionnaire « select fishing » |
| Probabilité d'aller à la pêche | Entre $\frac{2}{7}$ et $\frac{5}{7}$ pour les pêcheurs professionnels Entre $\frac{1}{7}$ et $\frac{3}{7}$ pour les pêcheurs "annexes" | Leenhardt, 2016 |
| Durée de la pêche | 4h pour les pêcheurs "annexes" 8h pour les pêcheurs professionnels | Questionnaire « select fishing », |
| Sélectivité / expérience | Nombre au hasard entre 0 et 1, sans unité | Attribuée au hasard pour chaque pêcheur |
| Période de pêche | Jour ou nuit | Attribuée au hasard pour chaque pêcheur, avec 70% de chance que ce soit la nuit, et 30% de chance le jour |
| Probabilité de braconnage | = 1 / (1 + surveillance) pour pêcheurs de nuit = dependance_ressource / (1 + surveillance) pour pêcheurs de jour | - |
| Taux de capture des poissons | Taux_capture_nuit = 2 * taux_capture_jour Avec taux_capture_jour = 0,002 $j^{-1}$ | Questionnaire « select fishing » |

Habitats préférentiels de pêche :

Les pêcheurs de Moorea ont des zones préférentielles de pêche. Dans le cadre des enquêtes sociales et des travaux menés par Thiault et al., 2017, un indice de préférence a été calculé spatialement.

Celui-ci dépend de 5 critères : distance à la côte, profondeur, distance à la passe, pente, substrat (corail, algue, sédiment). Afin d'évaluer le poids de ces différents critères sur la préférence, 52 pêcheurs de Moorea ont été interviewés.

Les différents critères ont été cartographiés à partir d'images satellite. En croisant les données satellitaires et les résultats des enquêtes, une carte de préférence moyenne des pêcheurs autour de l'île a été produite (figure 7). Ce critère de préférence sera utilisé ensuite pour influencer la zone de pêche du pêcheur.

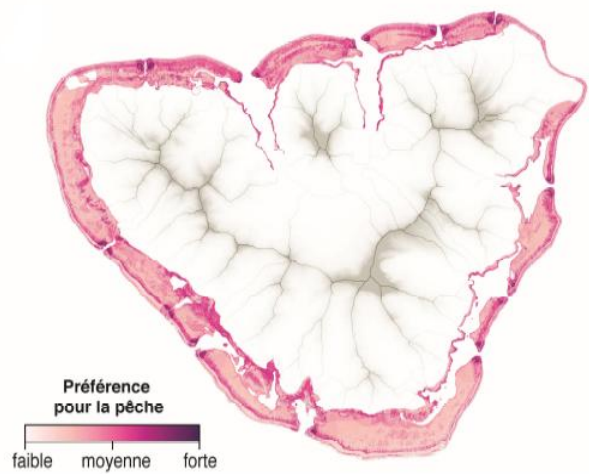

Figure 7 : carte de préférence de pêche
Source : Thiault et al., 2017

- Dynamique du pêcheur

  ◦ Zone de pêche :

Chaque pêcheur se voit attribuer une case au niveau de la côte qui correspond à la zone d'où il part pêcher. Nous avons choisi de prendre la case en mer la plus proche de son ménage. Nous l'appellerons case de départ du pêcheur.

Pour chaque pêcheur, nous lui attribuons une « zone potentielle de pêche ». Cette zone correspond à l'ensemble des cases où il est susceptible d'aller pêcher. Plus précisément, il s'agit des cases qui se trouvent à la distance « fishing radius » de la case de départ du pêcheur. Il ne s'agit pas d'une distance à vol d'oiseau, mais elle prend en compte les obstacles éventuels. Par exemple, pour aller en pente externe, les pêcheurs ne peuvent pas franchir le récif barrière et doivent donc faire un détour et passer par les passes.

Parmi ces cases « potentielles », le pêcheur va choisir une case que sera la zone de pêche, comme illustré dans la figure 8 :

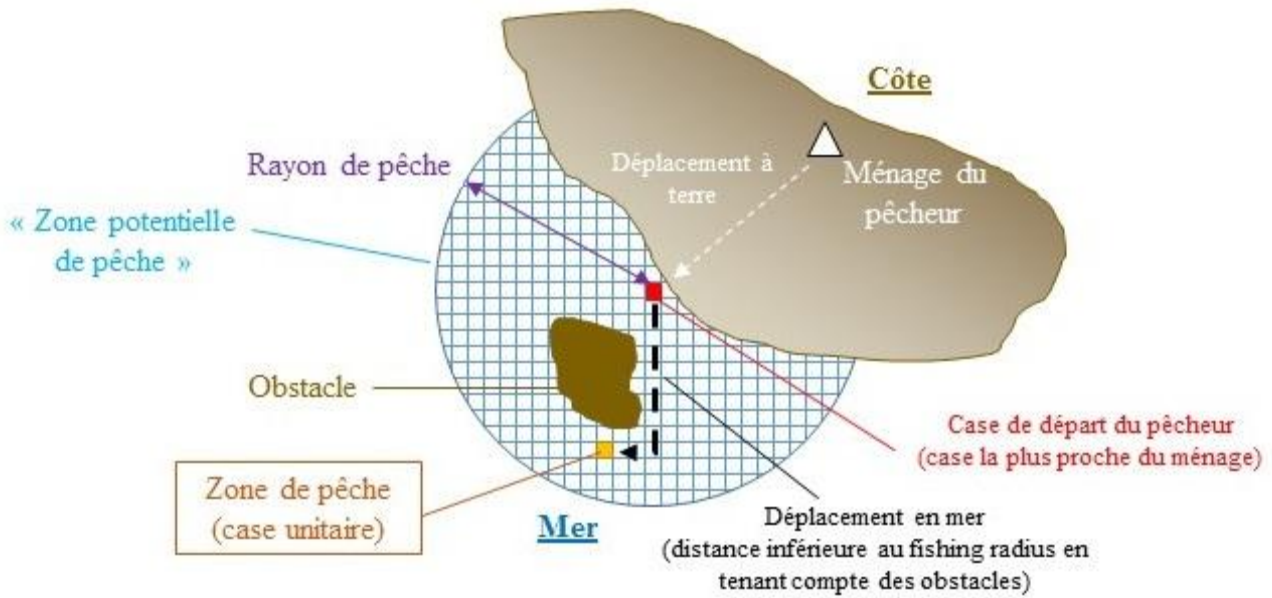

Figure 8 : Illustration de la « zone potentielle de pêche » et de la zone de pêche

Parmi les cases à l'intérieur de ce cercle, le pêcheur a une plus grande probabilité de choisir la case où la combinaison de la préférence de pêche, la biomasse de poissons, et l'inverse de la distance à parcourir est maximale, où dans une zone où il n'y a pas de pêcheur et une faible pression touristique.

$$\text{Preference}_{criterion} = \frac{\ln\left(\frac{preference}{preference\ max} + 1\right)}{\ln(2)} \quad ; \quad \text{Biomasse}_{criterion} = \frac{\ln\left(\frac{biomasse}{biomasse\ max} + 1\right)}{\ln(2)}$$

$$\text{Tourism}_{criterion} = 1 - \frac{\ln\left(\frac{level\ tourism}{tourisme\ max} + 1\right)}{\ln(2)} \quad ; \quad \text{Distance}_{criterion} = 1 - \frac{\ln\left(\frac{(distance\ à\ réaliser)}{(fishing\ radius)} + 1\right)}{\ln(2)}$$

Ces transformations ont pour but de lisser les valeurs, et d'obtenir des valeurs de critère entre 0 et 1.

Un pêcheur a moins de chance d'aller dans une case où un pêcheur est présent, par rapport à une case sans pêcheur. Nous attribuons un critère de pêcheur (pecheur $_{criterion}$) : il vaut 0.4 s'il n'y a pas de pêcheur ; 0.2 s'il y en a un ou plusieurs.

Ainsi, le pêcheur a une plus grande probabilité de choisir une case dans sa zone potentielle qui maximise la somme :

$$preference_{criterion} + biomasse_{criterion} + tourism_{criterion} + distance_{criterion} + pecheur_{criterion}$$

Si un autre pêcheur se trouve déjà dans la zone de pêche, le pêcheur se déplace dans une des cases voisines, préférentiellement dans une case sans pêcheur.
Quand le pêcheur ne pêche pas, sa localisation se situe au niveau de son ménage.

◦ Captures :

Biomasse capturée : une fois que le pêcheur a choisi une zone de pêche, il en extrait une certaine quantité d'herbivores et de carnivores. Les corallivores ne sont pas pêchés. Pour calculer la biomasse capturée par sortie de pêche, nous nous servons de la biomasse de poissons (herbivores + carnivores) présents dans la case, et de paramètres de pêche décrits précédemment :

$$Biomasse\ capturée = Biomasse\ poissons \times durée\ pêche \times (1 - sélectivité) \times taux_{capture}$$

◦ Conflits :

Un conflit se produit lorsqu'un pêcheur rencontre un autre pêcheur au niveau de sa zone de pêche, ou bien une zone touristique. De plus, lorsqu'un pêcheur va sur une case où il y a déjà un pêcheur, il choisit une autre case autour, préférentiellement là où il n'y a pas de pêcheur.

### II. 2. c) Tourisme

Dans le cadre des travaux réalisés par Charles Loiseau (avec Joachim Claudet) a été étudié la pression touristique en mer à Moorea du tourisme (notamment croisières / sorties bateaux).
L'indice de pression touristique, tel que représenté sur la figure 9, correspond au nombre de passages journaliers d'opérateurs touristiques dans une zone donnée. Ce sont des valeurs moyennes journalières.
A noter que l'on considère dans notre modèle que le tourisme est présent uniquement le jour, et non la nuit.

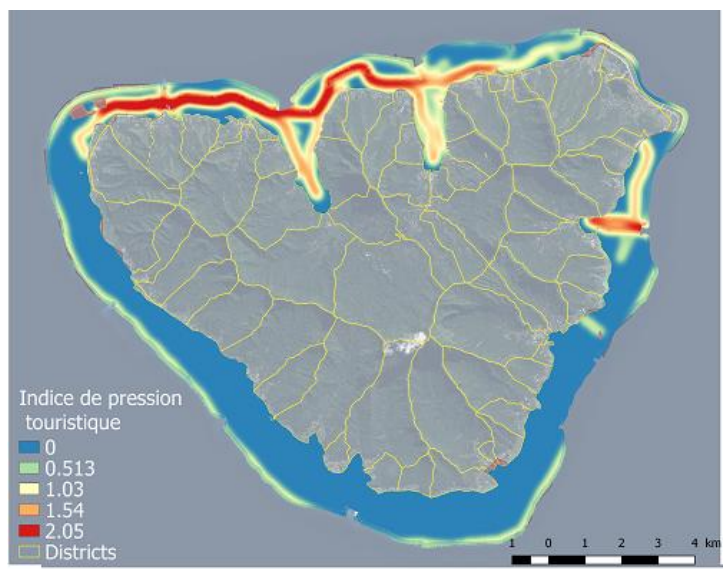

Figure 9 : indice de pression touristique autour de Moorea
*Source : Charles Loiseau, CRIOBE*

### II. 2. d) Calibrage des paramètres écologiques et conditions initiales

- Données spatialisées
  ◦ Répartition des recouvrement algues / corail

La carte de répartition du corail a été produite dans le cadre des travaux de *Collin A. et Hench J., 2015,* selon la méthode décrite en annexe 2. Les données datent de 2010, et nous obtenons des cartes de pourcentage de recouvrement de la surface.

Pour le turf, nous avons utilisé les données CRIOBE (*Data Substrate Cover*) qui fournit un recouvrement en algues dans le lagon et la pente externe entre 2005 et 2016, dont nous avons extrait une valeur temporelle moyenne pour le lagon et pour la pente externe.

Dans le tableau 3 sont présentées les moyennes spatiales obtenues :

Tableau 3 : moyenne spatiale des espèces benthiques, à l'état initial

|  | **Lagon** | **Pente externe** |
|---|---|---|
| **Corail** | 26 % | 44 % |
| **Turf** | 20 % | 33 % |
| **Macro-algues** | 3% | 3% |

◦ Répartition des biomasses de poissons

Pour les biomasses initiales de poissons (herbivores, corallivores, carnivores), nous avons utilisé les données *CRIOBE* (*Moorea Fish* - transects réalisés sur 13 sites différents autour de l'île, de 2004 à 2016) et *LTER* (*Annual Fish Surveys* - transects réalisés sur 6 sites différents, de 2005 à 2015).
Dans ces transects, sont recencés les espèces de poissons, leur nombre, et leur taille. Nous avons regroupé ces espèces dans nos 3 groupes (carnivores, corallivores, herbivores), nous avons calculé les biomasses par site et par groupe, et, grâce à un traitement statistique (logiciel R), nous avons pu obtenir la carte de répartition des biomasses de poissons autour de l'île. Nous avons utilisé et moyenné les données de 2006 et 2007, années pour lesquelles le système trophique était dans un état stable.

Dans le tableau 4 sont illustrés les biomasses moyennes (moyenne spatiale sur l'île et temporelle pour 2006 et 2007) que nous avons obtenues (par case de 10 000 m²) :

Tableau 4 : Biomasse moyenne des groupes trophiques, à l'état initial

| **Herbivores** | **Corallivores** | **Carnivores** |
|---|---|---|
| 305 kg | 25,3 kg | 78,5 kg |

- Paramètres biologiques

La paramétrisation s'est faite initialement à partir des paramètres biologiques estimés pour un modèle trophique (Dubois et al., 2017, en préparation) puis a dû être ajustée pour obtenir la dynamique observée dans les séries temporelles du CRIOBE (figure 4).

Pour les différents groupes d'agents, nous avons extrait le facteur $\frac{Production}{Biomasse}$ que nous avons assimilé au taux de croissance (et également au taux de mortalité d'après l'équilibre de Lotka Volterra).

Les taux de croissance α obtenus sont indiqués dans la tableau 5, et sont exprimés en jour$^{-1}$.

Tableau 5 : taux de croissance des agents du modèle

| **Corail** | **Turf** | **Herbivores** | **Corallivores** | **Carnivores** |
|---|---|---|---|---|
| $3.0 * 10^{-5}$ | $1.0 * 10^{-2}$ | $9.4 * 10^{-4}$ | $4.2 * 10^{-4}$ | $4.4 * 10^{-4}$ |

Dans les équations de Lotka-Volterra, les coraux sont influencés négativement par le corail (avec une taux de destruction coeff_destruction_corail_COTS), tandis que les carnivores et herbivores sont influencés négativement par les pêcheurs (avec un taux de mortalité $\beta_p$).

Pour le coefficient de destruction du corail par *Acanthaster plancii*, nous avons utilisé les données CRIOBE (*Model Substrate Cover*) d'évolution de la couverture corallienne en pente externe entre 2006 et 2010 (*Acanthaster plancii* étant arrivé en 2006).

On a :  $\text{coeff}_{\text{destruction\_COTS}} = \dfrac{\ln(\frac{c_{2010}}{c_{2006}})}{\Delta t} = 9{,}2 * 10^{-4}\ j^{-1}$

Le coefficient $\beta_p$ de mortalité par pêche est défini pour chaque case ; il vaut :

$$\beta_{pêcheur} = \dfrac{\alpha_{poisson}}{nombre\ de\ pêcheurs\ moyens\ par\ case}$$

Avec nombre de pêcheurs moyenne par case = $\dfrac{nombre\ de\ pêcheurs\ total}{nombre\ de\ cases} = \dfrac{2244}{5320} = 0{,}42$

Les autres paramètres (β, γ, δ) sont déterminés à partir des relations de Lotka-Volterra (2) à l'équilibre décrits dans la partie II. 2. a).

## *II. 3. Scénarios de gestion*

Dans le cadre de la révision du plan de gestion de l'espace maritime (PGEM), différents scénarios sont envisagés, et c'est eux que nous allons tester et comparer. Chaque scénario se traduit dans le modèle par la modification d'un paramètre social (tableau 6).

Tableau 6 : scénarios de gestion à tester

| Scénario sociologique | Conséquences sur les paramètres |
|---|---|
| Situation témoin / pas de gestion / statu quo | - |
| Pas de pêche (théorique) | 0 pêcheur dans le modèle |
| Respect théorique de la réglementation / absence de braconnage | Très faible probabilité de braconnage |
| Gestion à long terme : mise en place d'un quota journalier | Variable quota = 5 kg / jour (valeur moyenne proposée par les pêcheurs dans le questionnaire « select fishing »). Un pêcheur n'a pas le droit de dépasser ce quota |
| Aide financière aux pêcheurs pour l'acquisition de bateaux | Augmentation de la mobilité des pêcheurs. Nous prenons une aide financière de 50 %. Radius = radius * (1 + aide_financière) = radius * 1,5 |
| Interdiction de pêche la nuit | Les pêcheurs de nuit pêchent de jour à la place |

On simule chaque scénario sans perturbation et avec une perturbation environnementale.

## II. 4. Sorties du modèle et évaluation des scénarios

Dans un premier temps, l'objectif était de valider le modèle en comparant les sorties du modèle aux dynamiques observées à Moorea. Pour ce faire, différentes dynamiques temporelles ont été analysées :

- Dynamique du substrat en pente externe, correspondant à l'évolution du recouvrement moyen par case du corail et des algues (%) à chaque pas de temps
- Dynamique des poissons, soit l'évolution des biomasses moyennes par case (kg) d'herbivores, de carnivores et de corallivores à chaque pas de temps
- Dynamique des captures, soit l'évolution des biomasses pêchées par l'ensemble des pêcheurs de l'île (kg), à chaque pas de temps

Chaque scénario a ensuite été analysé à l'aide de 3 indicateurs :

- Indicateur de variation de biomasse de poissons dans l'eau :

$$Variation\ biomasse\ poissons = \frac{B_{final}}{B_{initial}}$$

Avec $B_{final}$ la biomasse de poissons par case à t = 20 ans (kg / ha) et $B_{initial}$ (kg / ha) la biomasse de poissons par case à t = 0 (kg / ha).

- Indicateur des captures annuelles moyennes :

$$Captures\ annuelles\ = \frac{\sum_{t=0}^{t=14000} Bp_t}{20}$$

Avec $Bp_t$ la biomasse totale pêchée à l'instant t.
14 000 correspond au nombre de pas de temps, et 20 au nombre d'années.

- Indicateur des conflits annuels moyens :

$$Conflits\ annuels\ = \frac{\sum_{t=0}^{t=14000} C_t}{20}$$

Avec $C_t$ le nombre de conflits à l'instant t

Ces indicateurs ont été calculés pour chaque case, et également au niveau global en faisant la moyenne des cases. Au niveau global, on regarde les captures annuelles et les conflits annuels par pêcheur, il faut donc diviser par le nombre de pêcheurs (soit 2244).

Afin de comparer l'effet de chaque scénario de gestion, les indicateurs ont été comparé aux valeurs obtenues dans le scénario « statu quo », comme suit :

$$Ratio = \frac{indicateur\ (scénario\ s)}{indicateur\ (statut\ quo)}$$

En comparant uniquement entre eux les scénarios ayant les même conditions environnementales (i.e. avec ou sans perturbation écologique).

Ci-dessous, nous présentons un résumé schématique des liens entre les paramètres de pêche, les scénarios, et les sorties du modèle (figure 10) :

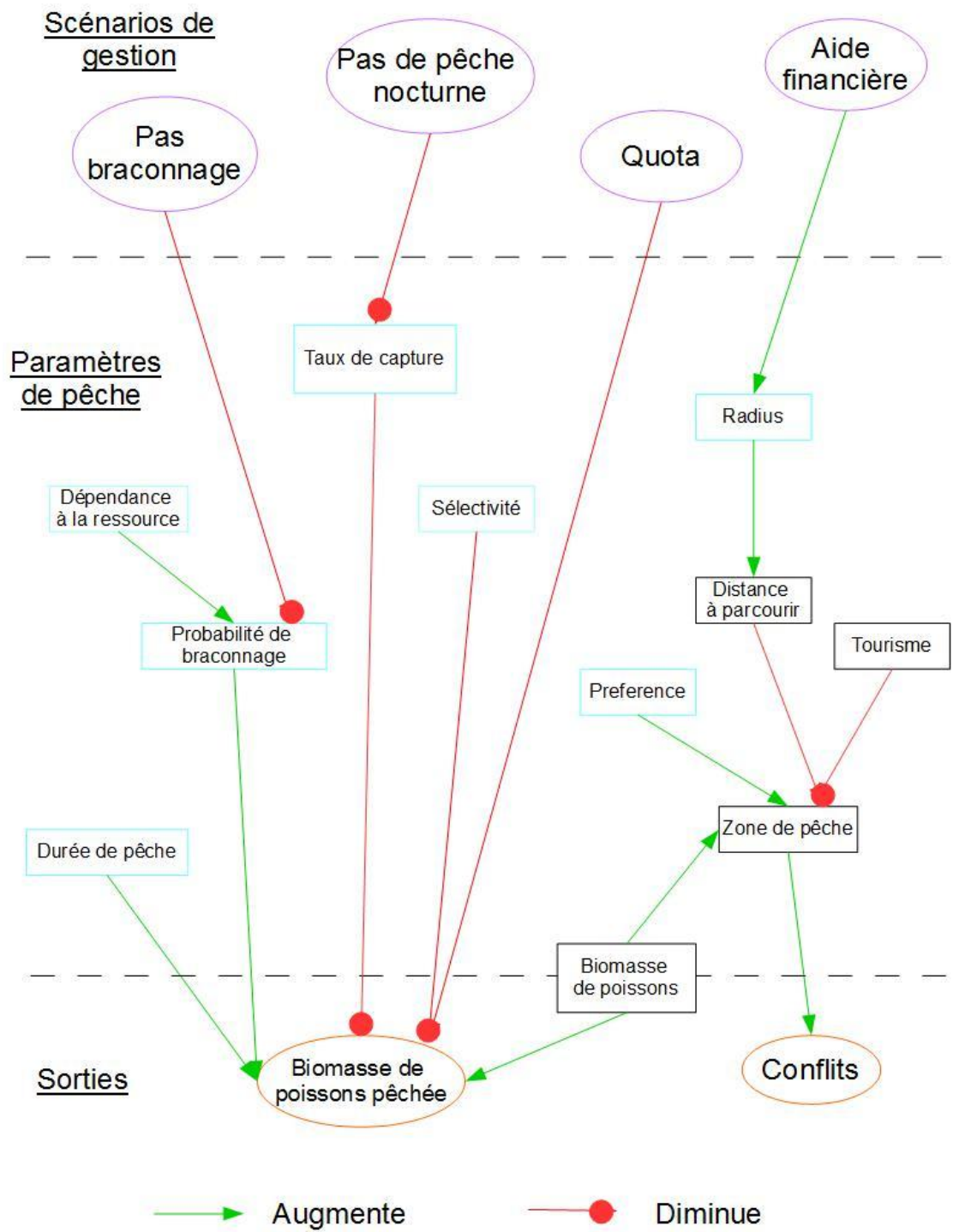

Figure 10 : Résumé des paramètres des pêcheurs, et liens avec les scénarios

# III. Résultats

## *III. 1. Validation du modèle*

Pour chaque simulation, la représentation des dynamiques temporelle de nos différents groupes écologiques a permis dans un premier temps de valider le comportement de notre modèle en les comparant à celles observées à Moorea. Le temps en abscisse est exprimé en nombre de pas de temps du modèle, soit le nombre de demi-jours (soit 14 000 demi-jours pour la durée de la simulation).

- Dynamique du substrat en pente externe en cas de perturbation environnementale (figure 11) :

On observe qu'une perturbation environnementale se traduit par une diminution du corail pendant 5 ans jusqu'à atteindre un minimum, puis une remontée jusqu'à atteindre l'état initial 10 ans après le début de la perturbation. Le turf, quant à lui, a une évolution inverse : il augmente jusqu'à 70% de recouvrement, puis diminue jusqu'à atteindre son niveau initial avant perturbation. Ceci est cohérent avec les séries temporelles présentées en figure 4.

- Biomasse de poissons dans l'eau : distinction AMP (aire marine protégée) / hors AMP (figure 12) :

En cas de non braconnage, c'est-à-dire de respect total de la réglementation, les herbivores et carnivores en AMP augmentent jusqu'à atteindre la capacité de charge (t = 11 ans pour les herbivores). Ceci est cohérent avec le scénario « pas de braconnage » dans lequel plus aucun pêcheur ne va pêcher dans une AMP.

- Biomasse de poissons dans l'eau : distinction lagon / pente externe (figure 13) :

Nous avons également simulé un scénario théorique sans pêcheur, et nous constatons que la biomasse en herbivores augmente bien et atteint la capacité de charge à t = 11 ans en pente externe et lagon.

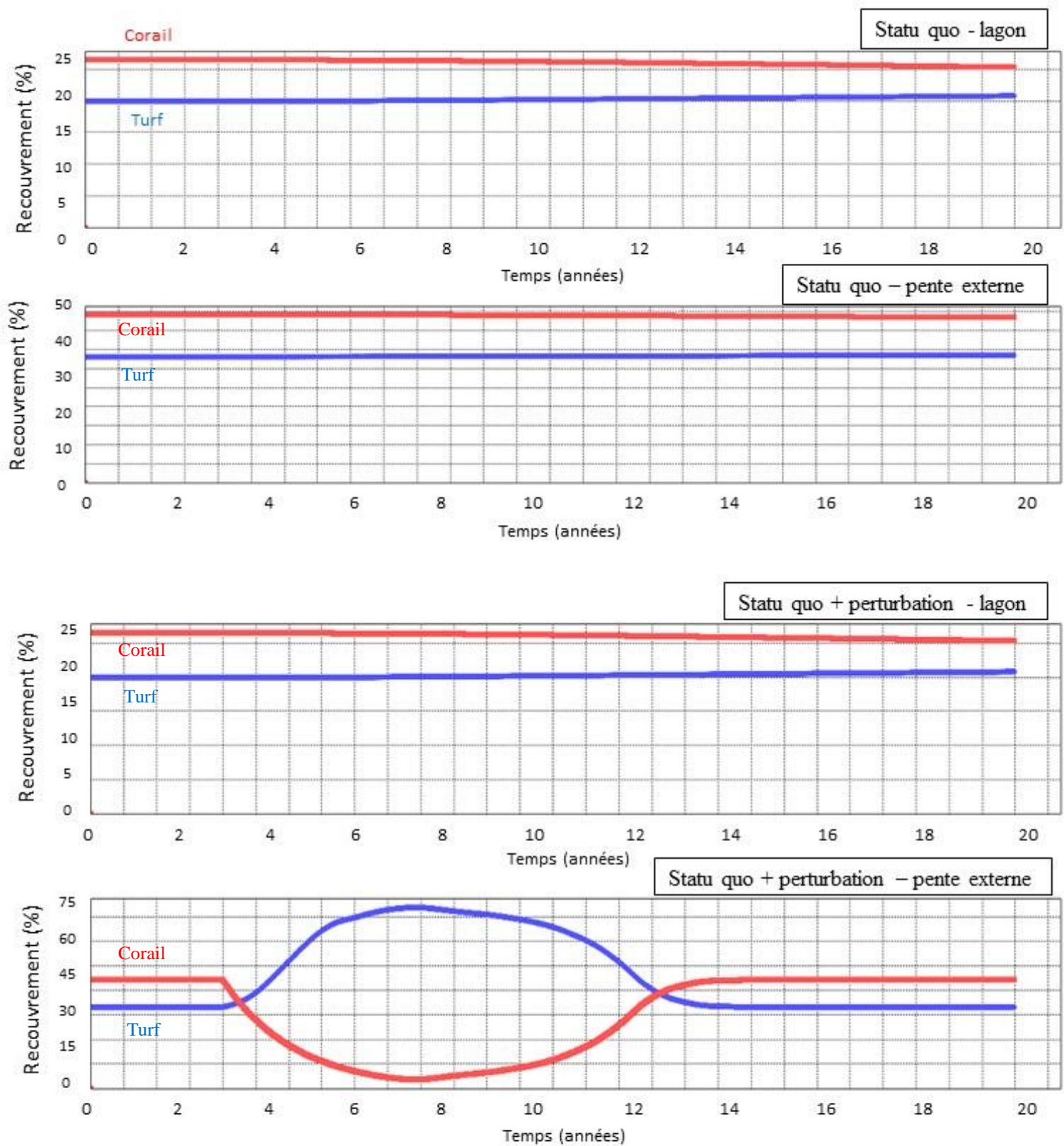

Figure 11 : Recouvrement moyen en corail et algues par case (en % de la surface) Distinction lagon / pente externe. Corail : courbe rouge ; turf : courbe bleue

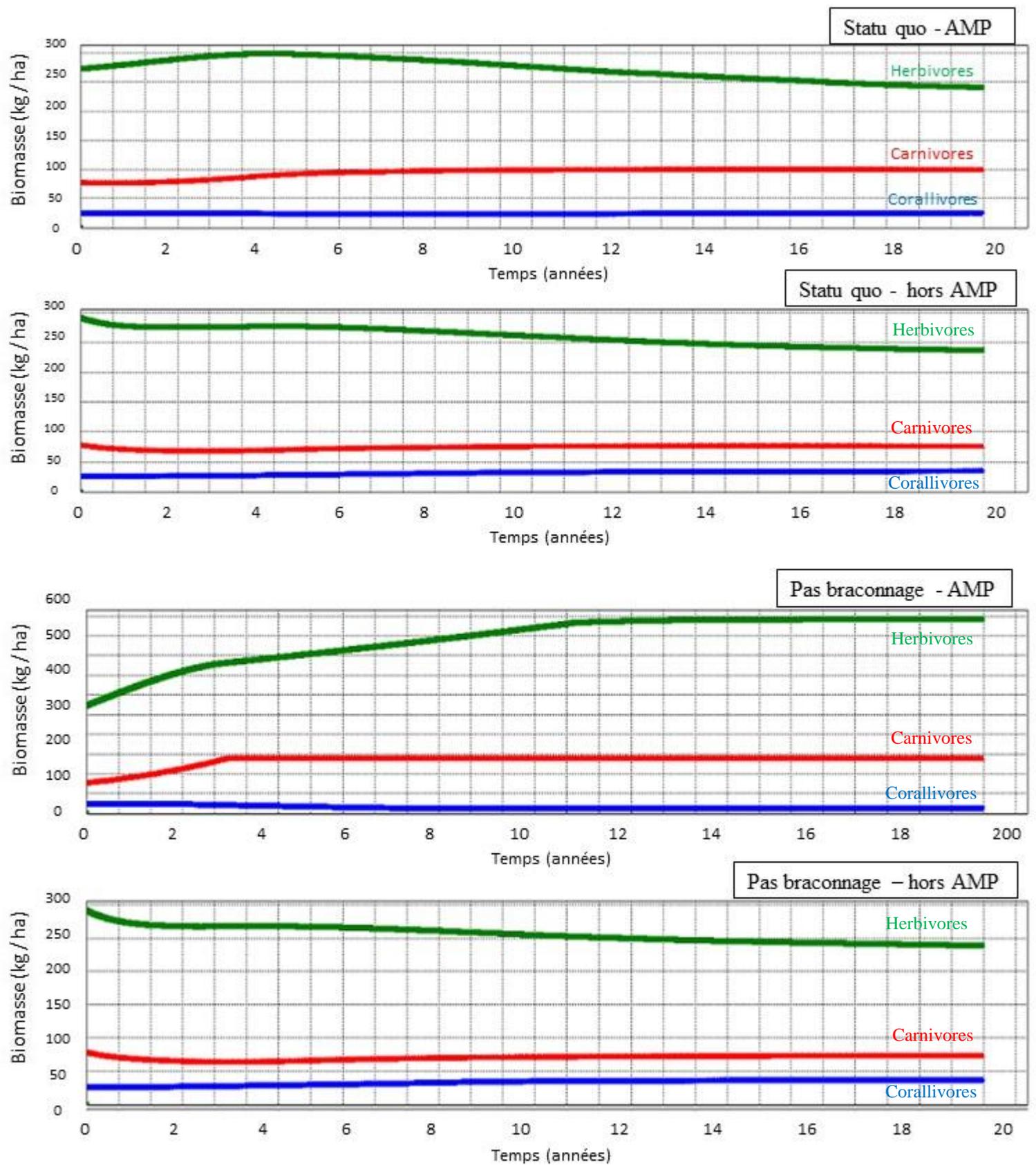

Figure 12 : Biomasse de poissons moyenne par case, à chaque pas de temps (kg)
Distinction AMP (aire marine protégée) / hors AMP
Herbivores en vert ; carnivores en rouge ; corallivores en bleu.

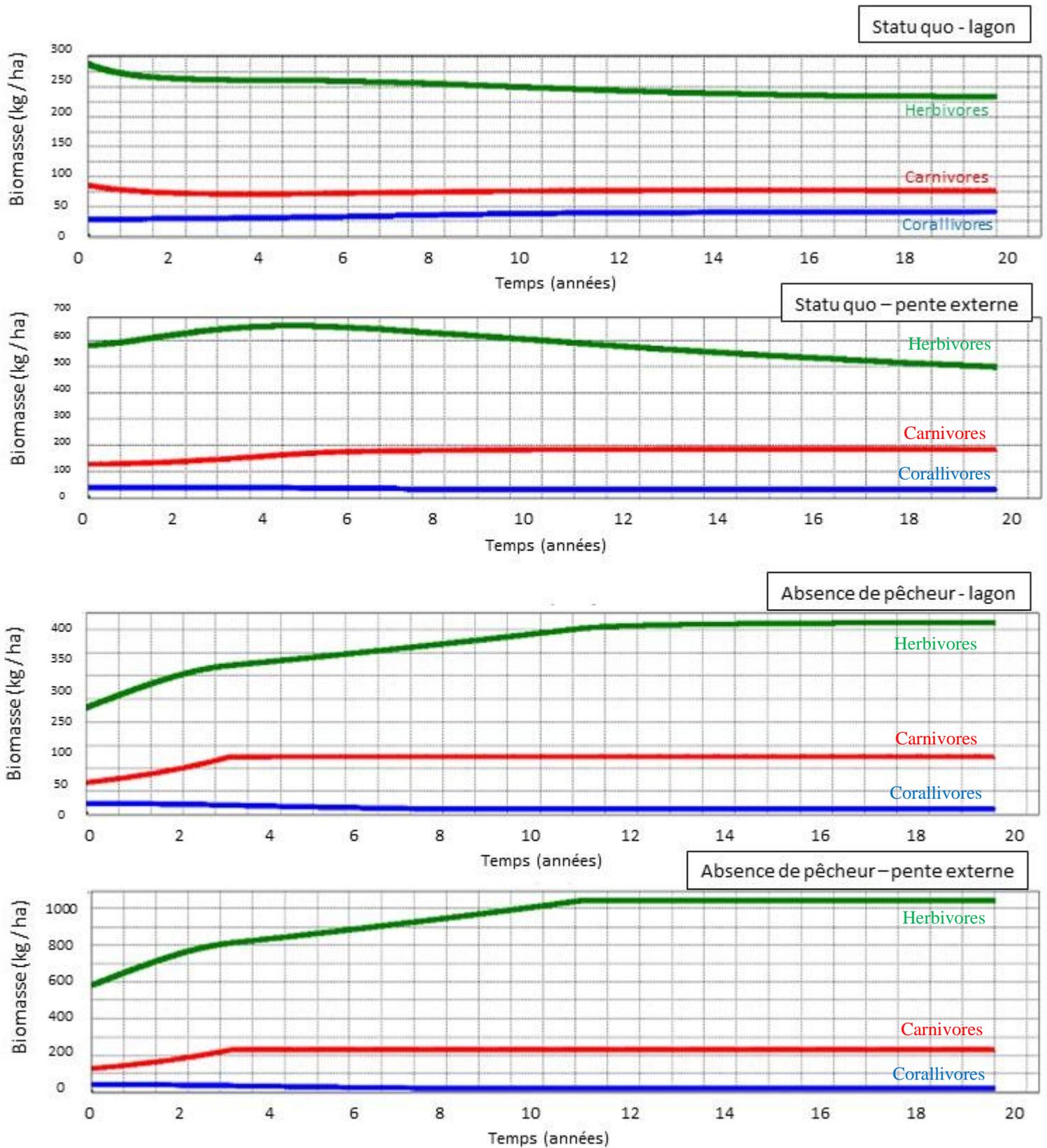

Figure 13 : Biomasse de poissons moyenne par case, à chaque pas de temps (kg)
Distinction lagon / pente externe.
Herbivores en vert ; carnivores en rouge ; corallivores en bleu

## III. 2. Analyse du scénario de référence – statu quo

Nous présentons dans le tableau 7 les indicateurs globaux - présentés dans la partie II. 4. – pour les scénarios de référence (« business as usual ») statu quo et statu quo avec perturbation.

Tableau 7 : valeurs des indicateurs globaux pour le statu quo

| Scénario | Variation de biomasse de poissons dans l'eau | Captures annuelles par pêcheur (kg / an) | Nombre de conflits annuels par pêcheur (an$^{-1}$) |
|---|---|---|---|
| Sans perturbation | 0,87 | 144,4 | 10,1 |
| Avec perturbation | 0,93 | 157,9 | 10,0 |

Les différences de valeur entre perturbation / pas perturbation s'expliquent par les liens de causalité suivants :

Perturbation environnementale -> moins de corail -> développement du turf -> augmentation des herbivores -> augmentation de la biomasse pêchée en herbivores.

Les résultats cartographiques permettent de mettre en évidence l'hétérogénéité spatiale de nos indicateurs. Les indicateurs par case, tels que décrits en partie II. 4. – sont représentées dans la figure 14 (pour le statu quo) et la figure 15 (statu quo avec perturbation). Pour les conflits, nous avons distingué les conflits qui se produisent le jour, et ceux qui se produisent la nuit.

On observe que les captures sont plus importantes au nord-est de l'île et au niveau des baies (figure 14. b.). En effet, elles correspondent à des zones étroites, et donc à plus forte pression de pêche par unité de surface. Les poissons dans ces zones se développent moins (figure 14. a.), et les conflits sociaux sont également plus importants (figure 14. c. et 14 d.). En cas de perturbation environnementale, c'est essentiellement la biomasse de poissons en pente externe dans l'eau qui est affectée : elle augmente en cas de perturbation (figure 15. a.).

Les indicateurs ont également été représentés par commune en figure 16 ; à Moorea, une commune est constituée de différents districts. L'analyse des cartes des résultats par commune montre que la pression de pêche n'est pas homogène sur l'île. Nous constatons également que la répartition captures (figures 16. a. et 16. c.) entre communes est similaire avec ou sans perturbation, de même pour les conflits (figure 16. b. et 16. d.).

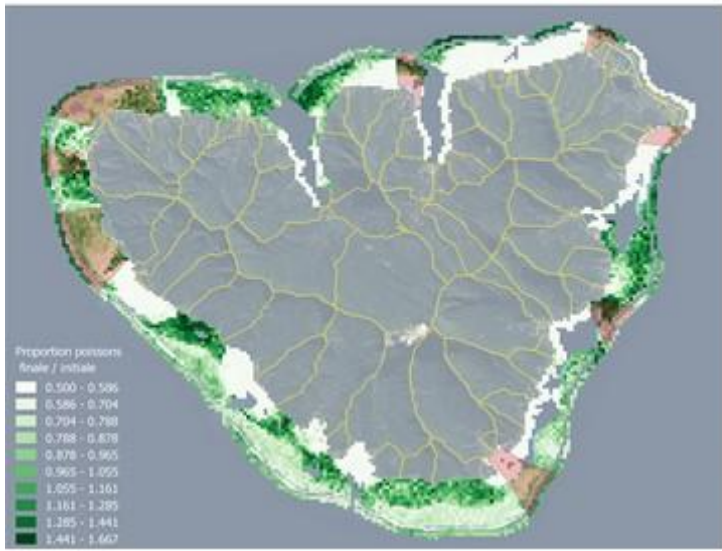

a) Variation de biomasse de poissons

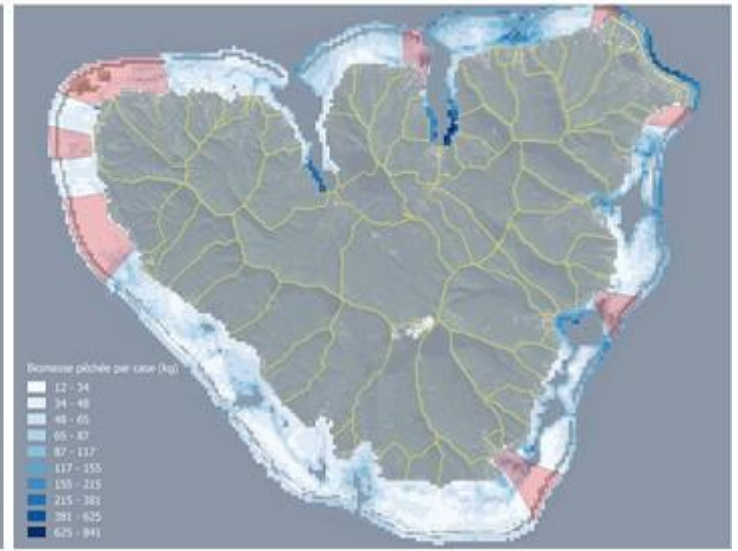

b) Captures annuelles (kg)

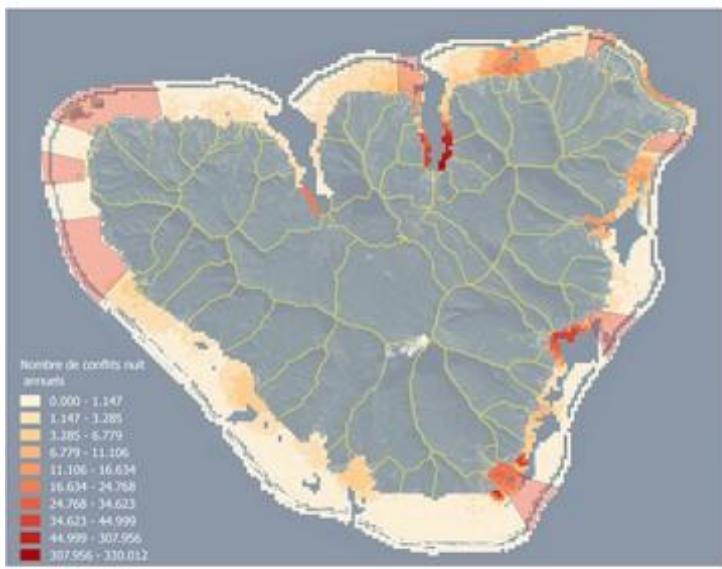

c) Nombre de conflits nocturnes annuels

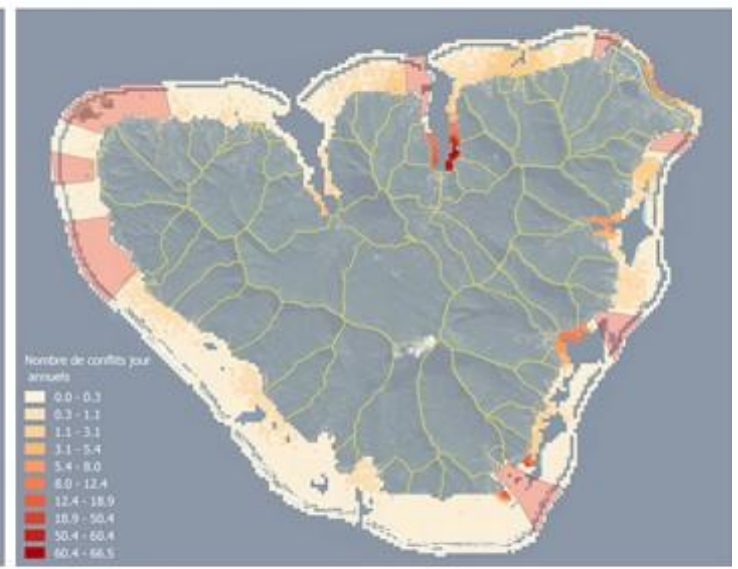

d) Nombre de conflits jour annuels

Figure 14 : Indicateurs par case pour le statu quo

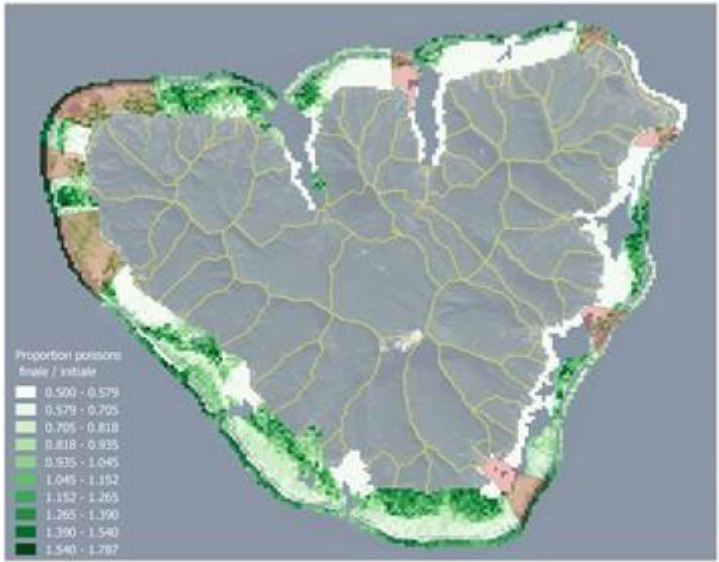
a) Variation de biomasse de poissons

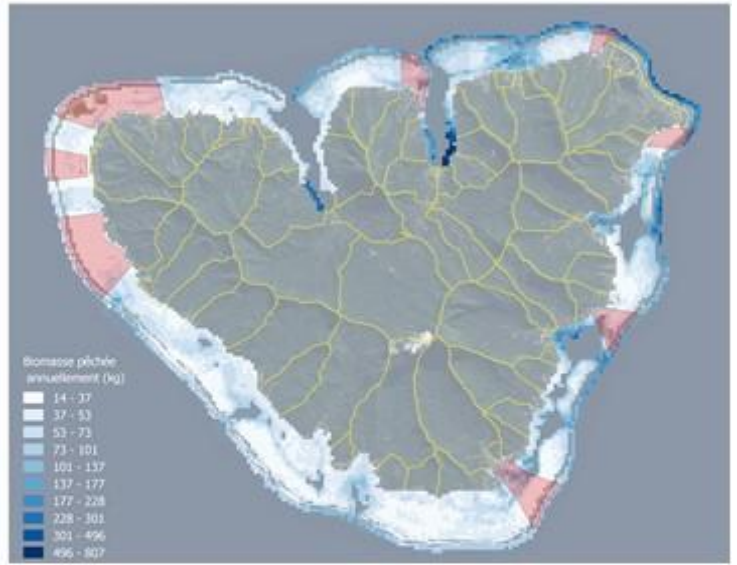
b) Captures annuelles (kg)

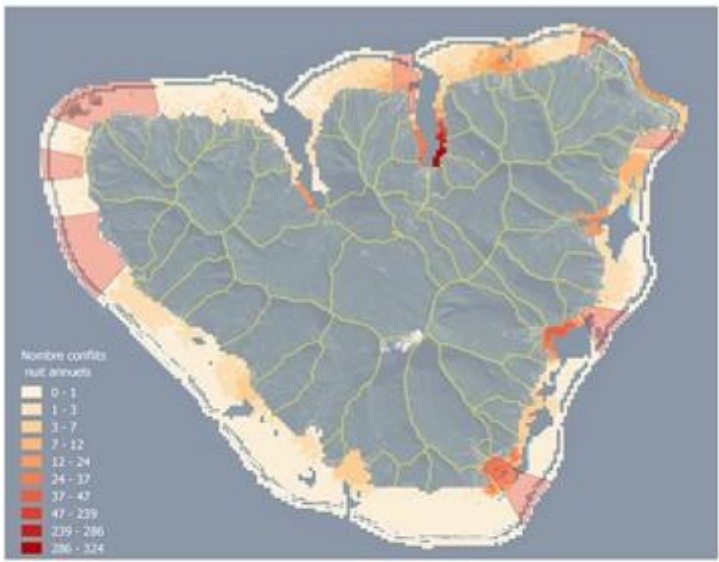
c) Nombre de conflits nocturnes annuels

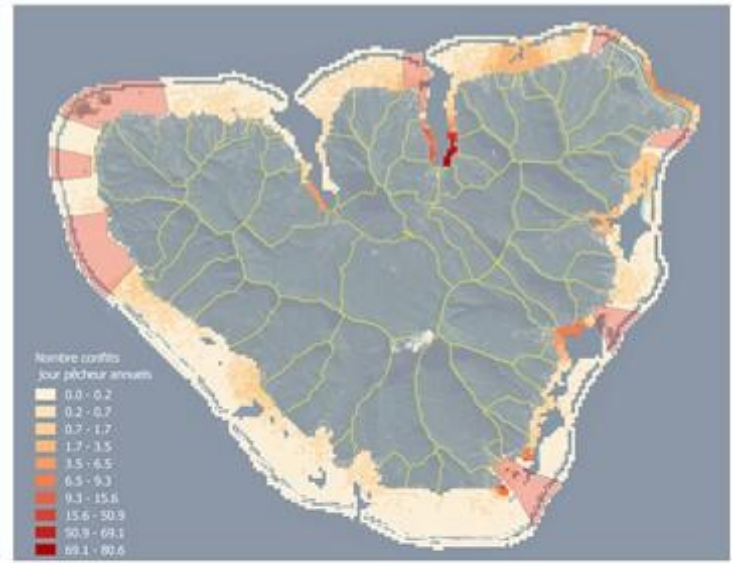
d) Nombre de conflits jour annuels

Figure 15 : Indicateurs par case pour le statu quo avec perturbation environnementale

Statu quo

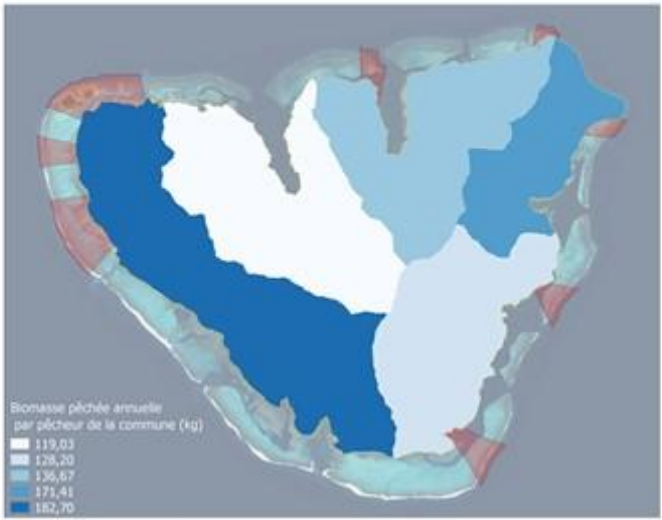

a) Captures annuelles par pêcheur de la commune (kg)

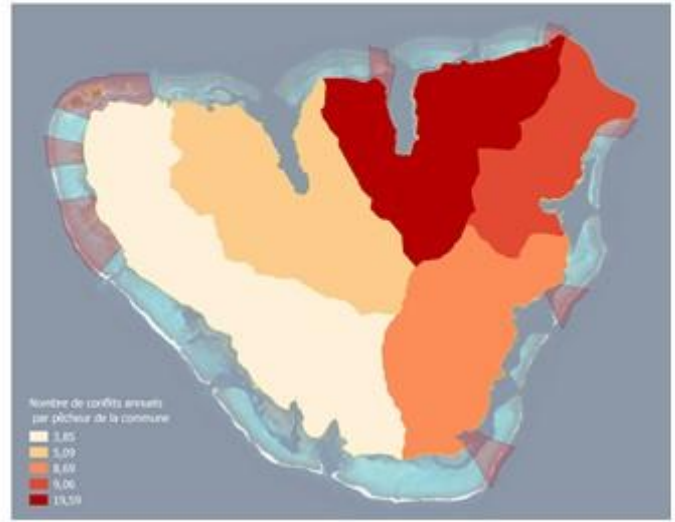

b) Conflits annuels par pêcheur de la commune

Statu quo + perturbation

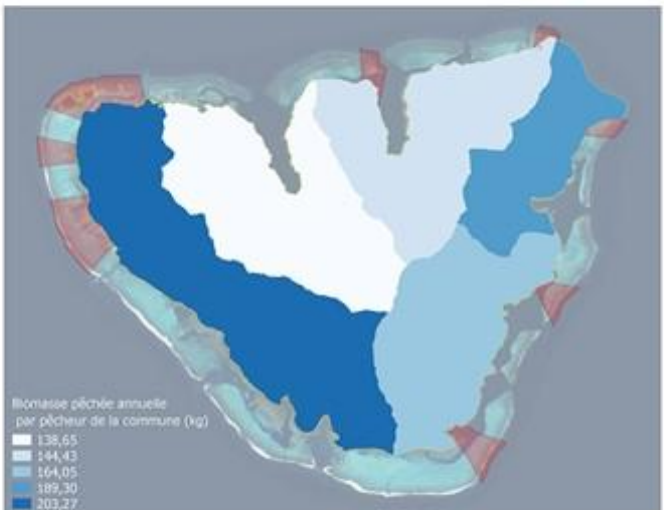

c) Captures annuelles par pêcheur de la commune (kg)

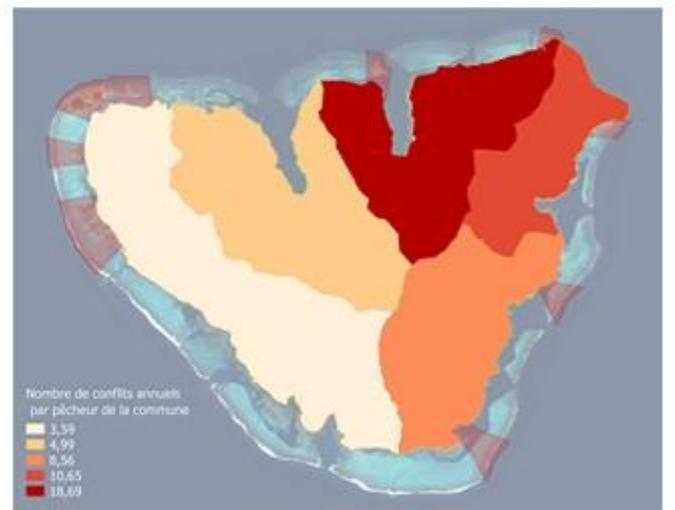

d) Conflits annuels par pêcheur de la commune

Figure 16 : Indicateurs par commune

## III. 3. Comparaison des scénarios

Nous représentons dans le tableau 8 les indicateurs pour les différents scénarios de gestion, en comparaison avec le statu quo : nous avons calculé les ratios tels que décrits en partie II. 4. Un ratio supérieur à 1 traduit une situation meilleure par rapport au statu quo (indiquée en vert dans le tableau) ; un ratio inférieur à 1 traduit une situation moins bénéfique (indiquée en rouge dans le tableau). En revanche, pour les conflits, la situation est inversée, car plus il y a de conflits, moins c'est favorable.

Tableau 8 : indicateurs globaux, en comparaison avec le statu quo

| Scénario de gestion | Perturbation | Variation de biomasse de poissons dans l'eau | Captures annuelles | Conflits annuels |
|---|---|---|---|---|
| Pas de braconnage | Non | 1,18 | 0,94 | 1,29 |
| | Oui | 1,13 | 0,92 | 1,24 |
| Quota | Non | 1,62 | 0,74 | 1,03 |
| | Oui | 1,54 | 0,69 | 1,07 |
| Interdiction pêche nuit | Non | 1,15 | 0,68 | 1,79 |
| | Oui | 1,17 | 0,66 | 1,79 |
| Aide financière | Non | 1 | 1,07 | 0,86 |
| | Oui | 0,98 | 1,06 | 0,83 |

Nous observons que les scénarios « pas braconnage », « quota » et « interdiction pêche nuit » ont des effets positifs sur la biomasse de poissons dans l'eau, alors que les indicateurs sociaux (captures et conflits) sont moins favorables par rapport au statu quo. Le scénario « aide financière », quant à lui, a des effets positifs sur les indicateurs sociaux (captures et nombre de conflits), sans impacter plus que le statu quo la biomasse de poissons dans l'eau.

## III. 4. Analyse spatiale des résultats

Afin d'analyser spatialement les résultats, nous avons produit des cartes qui représentent, pour chaque case, le ratio de l'indicateur i – tel que décrit dans la partie II. 4.

### III. 4. a) Indicateur de variation de biomasse de poissons

Les résultats cartographiques de l'indicateur « variation de biomasse de poissons dans l'eau » par case, en comparaison avec le statu quo, sont présentés en figure 17.

- Pour les scénarios sans braconnage (figure 17. b.) et interdiction pêche nuit (figure 17. c.), les quantités de poissons dans les aires marines protégées (AMPs) est significativement plus importante que dans les autres zones. En effet, en cas d'absence théorique de braconnage, les pêcheurs ne pêchent plus dans les AMPs ; en cas d'interdiction de pêche nocturne, la pêche s'effectue uniquement de jour et la probabilité de braconnage est alors plus faible (car il y a plus de surveillance).
De plus, nous observons des augmentations de biomasse de poissons autour des AMPs, ce qui traduit le phénomène de spillover à l'œuvre dans les AMPs.

- Pour les scénarios quota (figure 17. a.) et aide financière (figure 17. d.), nous n'observons pas de différence significative de biomasses dans l'eau entre les AMPs et les zones hors AMPs.

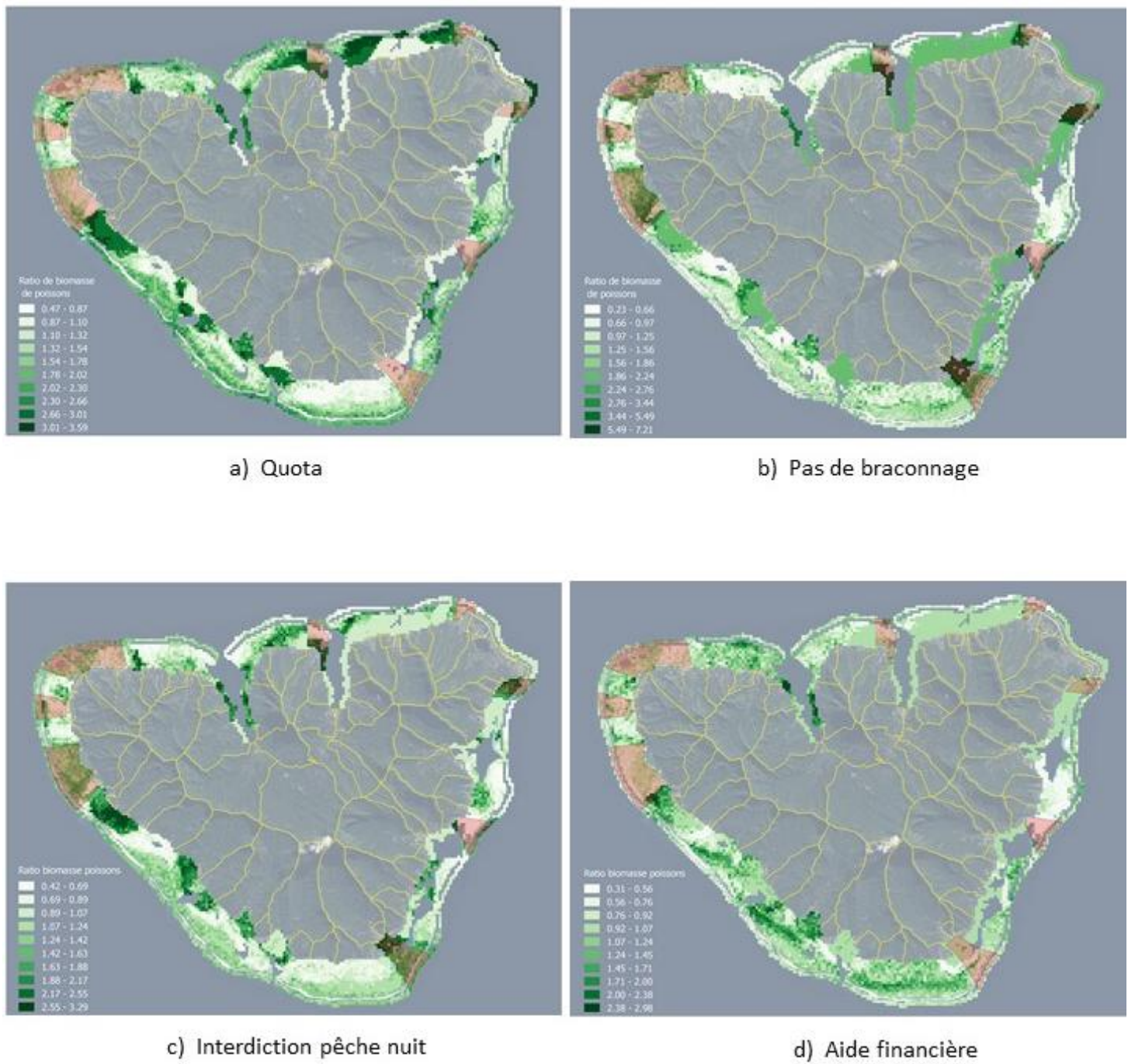

Figure 17 : Valeur par case de l'indicateur « variation de biomasse de poissons », en comparaison avec le statu quo

### III. 4. b) Indicateur des captures annuelles

Les résultats cartographiques de l'indicateur « captures annuelles moyennes » par case, en comparaison avec le statu quo, sont présentés en figure 18.

- Quota (figure 18. a.) : En cas de mise en place d'un quota, nous observons que la biomasse pêchée est plus importante près de la côte par rapport au statu quo.
- Absence théorique de braconnage (figure 18. b.) : la pression de pêche est évidemment nulle dans les AMPs, mais en dehors de ces zones elle est globalement plus importante par rapport au statu quo (et particulièrement à proximité des AMPs).
- Interdiction pêche nuit (figure 18.c.) : nous observons une situation assez similaire au quota. La différence principale est que la pression de pêche dans les AMP est nettement plus faible.
- Aide financière (figure 18. d.) : l'aide financière se traduit par une mobilité plus importante des pêcheurs, qui vont dès lors aller pêcher potentiellement plus loin. Nous voyons que la pression de pêche est plus importante en pente externe et à l'extérieur du lagon par rapport au statu quo.

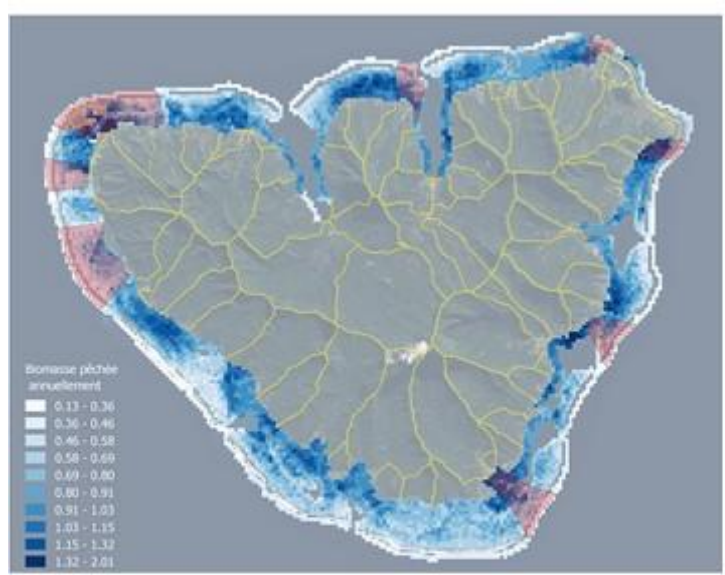

a) Quota

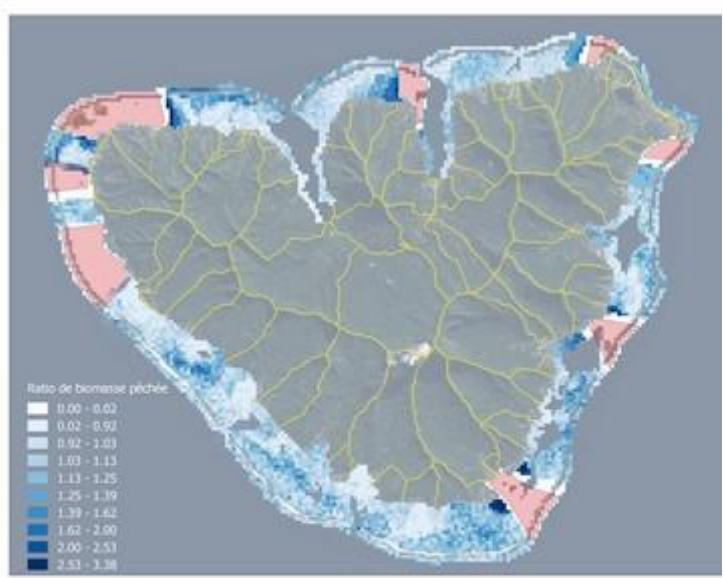

b) Pas de braconnage

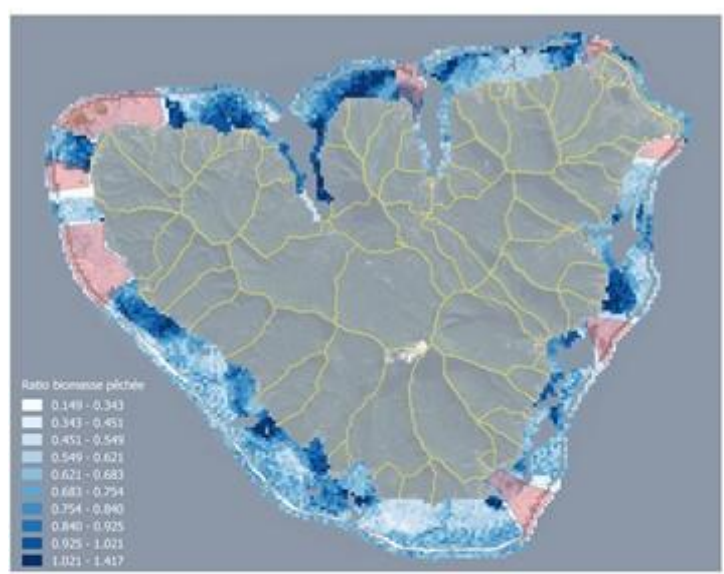

c) Interdiction pêche nuit

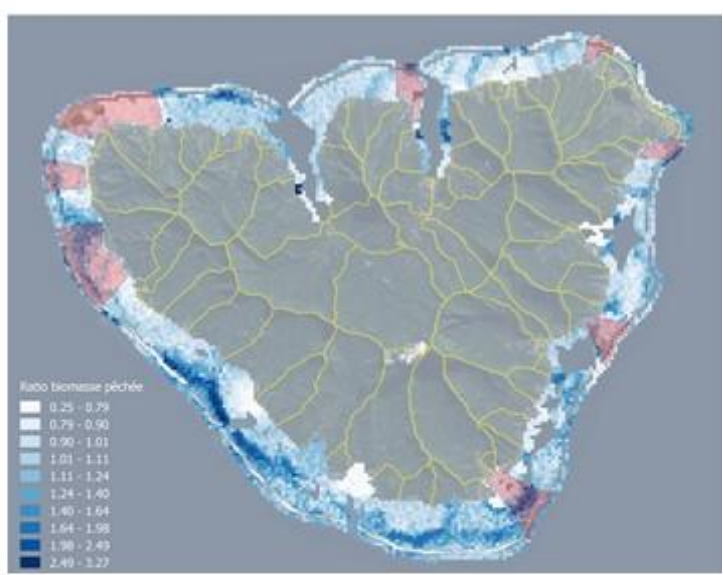

d) Aide financière

Figure 18 : valeur par case de l'indicateur « captures annuelles », en comparaison avec le statu quo

### III. 4. c) Indicateur des conflits annuels

Les résultats cartographiques de l'indicateur « conflits annuels moyens » par case, en comparaison avec le statu quo, sont présentés en figure 19.

- Pas braconnage (figure 19. b.) : nous observons que les conflits de nuit se situent principalement en pente externe et dans la partie externe du lagon.
- Quota (figure 19. a.) : nous n'observons pas de différence spatiale significative entre les conflits nocturnes dans le lagon et ceux en pente externe.
- Interdiction pêche nuit (figure 19. c.) : nous voyons sur la carte que les conflits jour sont globalement plus importants par rapport au statu quo, et ce notamment au niveau des baies et des zones lagonaires étroites.
- Aide financière (figure 19. d.) : les conflits sont plus importants en pente externe et dans la zone externe du lagon, ce qui est à corréler avec une pression de pêche plus importante dans ces zones.

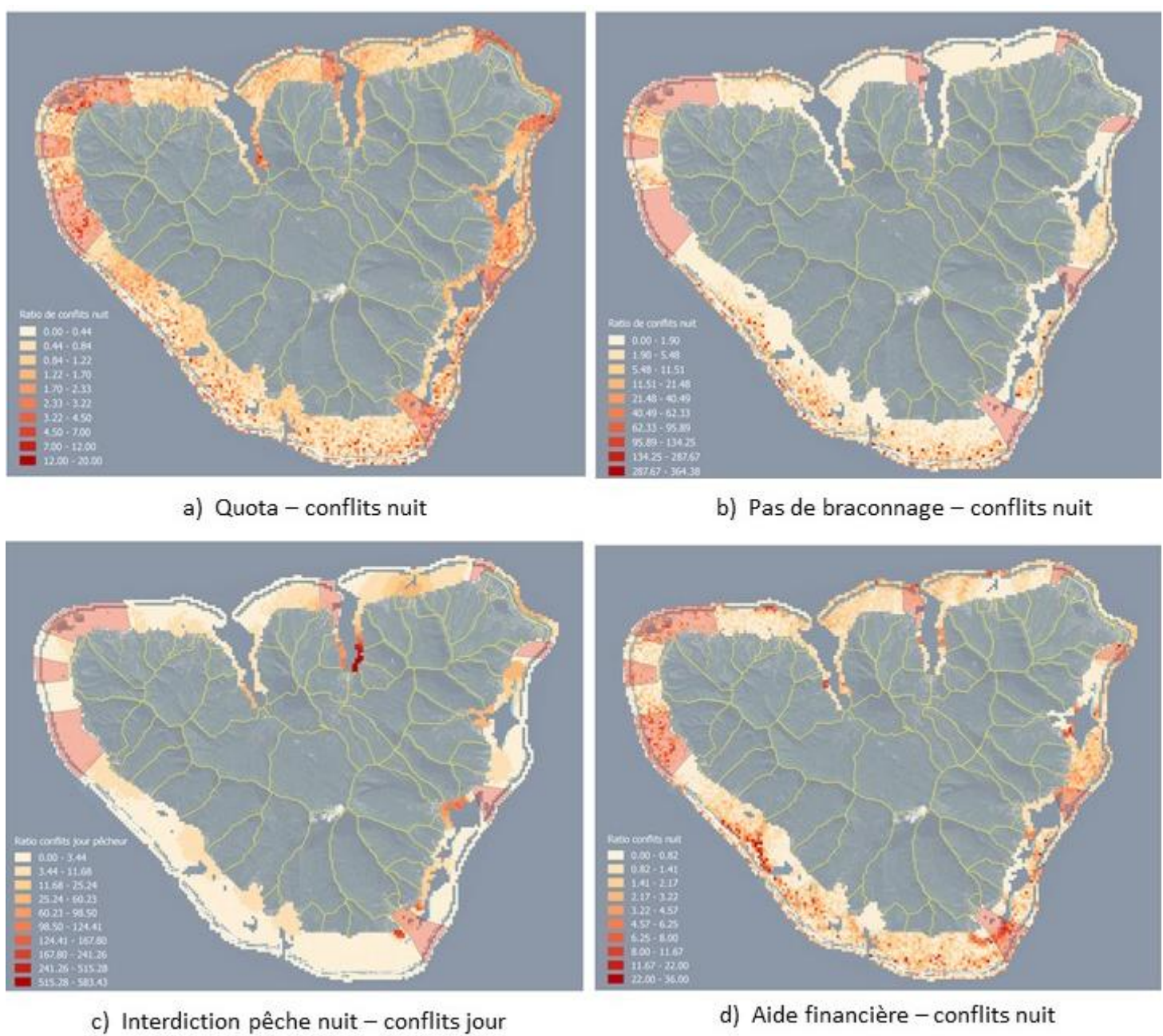

Figure 19 : valeur par case de l'indicateur « nombre de conflits annuels », en comparaison avec le statu quo

# IV. Discussion et perspectives

Analyse des scénarios : Le modèle multi-agents a permis une comparaison des scénarios de gestion testés, avec des effets observés sur les outputs désirés, et a fait ressortir des effets indirects et inattendus de ces scénarios.

Avec la mise en place du quota, il est moins avantageux d'aller loin car, bien qu'il y ait plus de ressource dans la pente externe que dans le lagon, ils ne pourront pas pêcher plus que le seuil autorisé. Dès lors, ils favorisent les zones à faible distance de la côte, ce qui se traduit par une pression de pêche plus importante dans ces zones. La mise en place d'un quota journalier bénéficie à la ressource, bien que la biomasse pêchée soit inférieure au statu quo. Mais sur le long terme, une quantité de poissons plus importante pourrait se traduire par une pêche plus importante par rapport au statu quo. La mise en place de quotas individuels est considérée de plus en plus comme un moyen de rendre les pêches plus rentables (Little et al., 2009).

L'absence théorique de braconnage créé un déplacement de l'effort de pêche des aires marines protégées (AMPs) vers les zones hors AMP, ce qui assure un bon renouvellement de la ressource dans les AMPs. Pour tendre vers cet état sans braconnage, des discussions sont nécessaires entre autorités, scientifiques et pêcheurs pour bien faire comprendre à ces derniers la légitimité d'une telle mesure. En effet, à travers le monde, beaucoup de mesures de conservations (dont aires marines) sont imposées sans prendre en compte les communautés locales, qui sont pourtant les premières touchées par ces décisions (Boissière et al., 2008).

Le jour, il y a plus de surveillance de la part des autorités, les pêcheurs braconnent donc moins. En cas d'interdiction de pêche nocturne, il y aura donc moins de braconnage dans les AMPs, ce qui se traduira par un meilleur état des ressources halieutiques dans ces zones. Cependant, interdire aux gens de pêcher la nuit risque d'être difficilement acceptable culturellement et conduirait à une augmentation des conflits entre pêcheurs et avec les opérateurs touristiques.

Enfin, l'aide financière se traduit par une situation en quelque sorte inversée par rapport au quota, où la pression de pêche et les conflits sont plus importants en pente externe. Ce résultat spatial indique que ce scénario est potentiellement plus sensible au scénario « quota » en cas de perturbation environnementale qui a lieu uniquement en pente externe (*Acanthaster plancii* ou cyclone). L'aide financière est une bonne solution pour mieux répartir la pression de pêche et éviter qu'il y ait des zones surexploitées. Cependant, comme toute aide financière, il faudrait trouver une méthode équitable de distribution de la ressource monétaire, ce qui pourrait occasionner des conflits à ce niveau (Hilborn, 2007).

A travers le monde, les avis divergent au sujet des modes de gestion, certains prônant la mise en place d'aires marines protégées, d'autres souhaitant une aide financière équitablement distribuée associée à une plus large participation des pêcheurs dans les décisions de gestion (Fulton et al., 2011). Quel que soit le mode de gestion choisi par les autorités, il ne pourra être mis en place qu'avec une concertation avec les pêcheurs, afin de leur expliquer la légitimité d'une telle mesure, les effets bénéfiques sur les écosystèmes, et ce qu'ils ont à gagner sur court et moyen termes.

Les scénarios testés ici ont eu peu d'influence sur la régénération du corail et sur la sensibilité du système à « shifter » vers une phase dominée par les algues. Il semblerait que la pêche ait peu d'influence sur cette sensibilité, à la différence des phénomènes liés au changement climatique (augmentation de la température des eaux, acidification, blanchissement du corail), et d'autres facteurs anthropiques (pollution des eaux aux nitrates par ex. qui favorisent le développement des algues) - Graham et al., 2013. La synergie de ces deux types de perturbation, environnementale et anthropique, ne facilite pas la récupération du corail. A travers le monde, les inquiétudes sont grandissantes quant à l'évolution des récifs coralliens, y compris à Moorea, dans un contexte de changement climatique et de pressions anthropiques de plus en plus importantes (Graham et al., 2013, Baskett et al., 2010).

L'originalité de cette étude est de considérer les deux aspects sociaux et écologiques à la fois, et d'analyser les répercussions de modes de gestion diverses sur la durabilité des ressources halieutique, la résilience de l'écosystème, et les bénéfices pour les pêcheurs. A notre connaissance, il s'agit de la première simulation multi-agents réalisée en Polynésie (Ghazia et al., 2014). Une simulation multi-agents a été menée sur la côte ouest de l'Australie, dans le Parc Marin de Ningaloo (Gao et al., 2012), en prenant en compte également le système socio-écologique corail et pêche. Les chercheurs ont testé l'effet de scénarios de fermeture de zones à la pêche pendant une certaine durée (2 ou 6 mois), associé à une variation de la pression de pêche, et ont considéré en sorties des critères sociaux, écologiques et économiques. Leur étude montre que l'efficacité d'une mesure d'AMP dépend fortement de la pression de pêche. En cas de faible pression de pêche, les bénéfices sociaux et écologiques sont plus importants quand des fermetures de zone de pêche sont mises en place pendant de longues périodes ; cette corrélation positive n'est plus valable en cas de pression de pêche importante.

Perspectives : L'adaptabilité du modèle à d'autres situations à Moorea, sous d'autres contextes économiques et sociaux, est réalisable en changeant les paramètres d'entrée.

Il serait opportun de mélanger les scénarios testés ici, par exemple quota + absence théorique de braconnage ; en effet, selon les travaux de Doyen et al., 2007 et Goa et al., 2011, une gestion viable à travers les AMPs requiert également un effort de régulation de la pression de pêche. Sur les 20 ans, nous avons vu un effet positif de l'aide financière ; cependant, de telles aides ont eu un effet néfaste à travers le monde sur les systèmes pêcheurs-poissons de par l'augmentation de la capacité engendrée (Milazzo, 1998). Il serait bon d'envisager de tester la viabilité de cette aide sur un temps plus long. Dans notre modèle, les outputs sociaux relatifs aux pêcheurs sont les conflits et la biomasse pêchée ; or, pour beaucoup de pêcheurs qui pratiquent la pêche de loisir, la biomasse pêchée importe peu, le simple fait de pouvoir aller à la pêche est une source de satisfaction. Il serait intéressant de prendre en compte ces bénéfices non-matériels, bien qu'ils soient difficiles à évaluer quantitativement (Leenhardt et al., 2016).

Le modèle est transposable à d'autres zones d'étude insulaires, en modifiant les données d'entrée de cartographie. Par exemple, il serait très intéressant de lancer le modèle pour une zone insulaire où les récifs coralliens ont perdu leur capacité de résilience, de comparer avec Moorea et de voir quels scénarios de gestion sont les plus favorables pour la résilience de l'écosystème. Le modèle multi-agents peut également être utilisé pour aider à informer les gestionnaires et décideurs dans leurs activités journalières, avec une actualisation des données au jour-le-jour, et ainsi éviter des phénomènes imprévus (Pereira et al. 2009).

Limites du modèle : Les systèmes biologiques et sociologiques sont régis par des dynamiques chaotiques, avec de multiples interactions, et sont de ce fait difficilement prévisibles (Ferreira et al. 2017). Afin de limiter ces interactions qui augmentent la complexité du système et les temps de calcul, nous avons considéré un nombre restreint d'agents, uniquement ceux en relation directe avec le corail et les pêcheurs. Les agents qui entretiennent des relations indirectes (plancton, bactéries par ex. au niveau écologique, agriculteurs au niveau social) n'ont pas été pris en compte, bien qu'ils exercent une influence sur le système. Spatialement, nous avons considéré un système fermé, Moorea, sans interaction avec l'extérieur, afin de simplifier l'étude. L'incertitude inhérente à la pêche à Moorea vient de la diversité des motivations de pêche, des moyens de locomotion, des périodes de pêche et des poissons pêchés (Leenhardt et al., 2016).

# Bibliographie

# Annexes

[1]   Equations de Lotka Volterra par case :

$$\begin{cases} \dfrac{dC}{dt} = C * (\alpha_C - \beta_T * T - \beta_{CO} * CO - coeff_{destruction_{COTS}} * COTS) \\[1em] \dfrac{dT}{dt} = T * (\alpha_T - \beta_H * H - \beta_c * C) \\[1em] \dfrac{dH}{dt} = H * (\alpha_H - Y_H - \beta_P * P - \beta_{pêche} * nb\_pêcheurs + \delta_T * T) \\[1em] \dfrac{dCO}{dt} = CO * (\alpha_{CO} - Y_{co} - \beta_P * P + \delta_c * c) \\[1em] \dfrac{dP}{dt} = P * (\alpha_p - Y_P - \beta_{pêche} * nb\_pêcheurs + \delta_{H-CO} * (H + CO)) \end{cases}$$

[2] : <u>Méthode d'obtention des cartes de répartition des algues et des coraux</u>

A partir des données satellitaires, sont extraits la profondeur, la composition de l'habitat à une précision de 5m. x 5m. au niveau du lagon et de la pente externe. En complément de ces données satellitaires, des relevés ont été réalisés entre juin 2010 et novembre 2011 à l'aide d'un bateau adapté qui prend des images des fonds marins.

En croisant ces deux sources de données, nous en retirons des probabilités de trouver du corail, des algues ou des sédiments aux différents endroits autour de Moorea.

Source : Collin et al., 2015

[3] : <u>Questionnaire de l'ISPF, bulletin individuel, 2007</u>

*[Formulaire de recensement de la population - 2007, Polynésie française, Bulletin individuel. Le questionnaire comprend les sections suivantes :]*

**1** Sexe : Masculin / Féminin

**2** Quelle est votre date de naissance ? (jour, mois, année) ; Sinon, âge en années

**3** Où êtes-vous né(e) ?
- En Polynésie française
  - Où (commune)
- En Métropole
- En Nouvelle-Calédonie ou à Wallis et Futuna
- Dans une autre collectivité d'outre-mer ou dom
- À l'étranger
  - Précisez le pays

**4** Quelle est votre nationalité ?
- Française
  - Vous êtes né(e) français(e)
  - Vous êtes devenu(e) français(e) (par naturalisation, mariage, déclaration ou à votre majorité)
    - Indiquez votre nationalité à la naissance
- Étrangère
  - Indiquez votre nationalité

**5** Si vous êtes né(e) en dehors de la Polynésie française, en quelle année vous y êtes-vous installé(e) pour la dernière fois ?

**6** Où habitiez-vous le 1er novembre 2002 ? (les enfants nés après cette date ne sont pas concernés)
- Dans le même logement
- En Polynésie française, mais dans un autre logement
  - À (commune)
  - (Commune associée, île)
- En métropole
- En Nouvelle-Calédonie ou à Wallis et Futuna
- Dans une autre collectivité d'outre-mer ou dom
- À l'étranger
  - Précisez le pays

Pour une personne de passage, adresse de sa résidence habituelle :
N°, PK, route, rue ; Commune ou commune associée (en clair)

*La suite du questionnaire s'adresse aux personnes de 14 ans ou plus. (personnes nées avant le 20 août 1993)*

**7** Vivez-vous actuellement en couple : Oui / Non

**8** Quel est votre état matrimonial ? Indiquez votre situation juridique présente
- Célibataire (jamais légalement marié(e))
- Marié(e)
- Veuf, veuve
- Divorcé(e)

**9** Pour les femmes uniquement
- Combien avez-vous eu d'enfants nés vivants ?
- Avez-vous donné des enfants à fa'a'amu ? Non / Oui — Si oui, combien ?
- Avez-vous actuellement des enfants en fa'a'amu ? Non / Oui — Si oui, combien ?

**10** Êtes-vous actuellement écolier, étudiant ou stagiaire rémunéré ? (année scolaire ou universitaire 2007-2008)
- Oui — Nom de l'établissement d'enseignement ou de stage
  - Commune ou pays si hors Polynésie française
- Non

**11** Connaissance des langues (Le français / Une langue polynésienne — Oui/Non)
- Vous comprenez
- Vous parlez
- Vous lisez
- Vous écrivez
- Quelle est la langue la plus couramment parlée en famille ?

**12** Quel est votre niveau d'études ?
- Aucune scolarité
- Scolarité jusqu'en école primaire
- Scolarité secondaire, jusqu'au collège (de la 6e à la 3e incluse)
- Scolarité secondaire, niveau BEP, CAP ou équivalent
- Scolarité secondaire (seconde, première, terminale)
- Études supérieures au-delà du baccalauréat

Quartier, lieu-dit, lotissement :

Continuez page suivante et n'oubliez pas de faire signer →

**13** Quel est le diplôme le plus élevé que vous avez obtenu ?
- Aucun diplôme
- CEP (certificat d'études primaires)
- BEPC, brevet élémentaire, brevet des collèges
- CAP, brevet de compagnon
- BEP
- Baccalauréat général, brevet supérieur
- Baccalauréat technologique ou professionnel, brevet professionnel ou de technicien, capacité en droit
- Diplôme de 1er cycle universitaire, BTS, DUT, diplôme des professions sociales ou de la santé, d'infirmier(ère)
- Diplôme de 2e cycle universitaire (licence, maîtrise ou équivalent)
- Diplôme de 3e cycle universitaire (y compris médecine, pharmacie, dentaire), diplôme d'ingénieur, d'une grande école, doctorat

**14** Quelle a été votre situation au cours de la semaine précédant l'interview ?
- Vous avez travaillé ou été absent(e) de votre travail pour des raisons temporaires (congé, maladie, maternité, etc.). (Cochez la case 1 également dans les deux cas suivants : la personne aide un membre de sa famille dans son travail ; la personne est un stagiaire rémunéré ou un apprenti sous contrat) → Allez à la question **18**
- Vous exercez une activité intermittente ou saisonnière mais vous n'avez pas travaillé pendant la semaine précédant l'interview. → Allez à la question **18**
- Élève, étudiant, stagiaire non rémunéré
- Sans travail et recherche d'un travail salarié
- Sans travail, en cours de création d'entreprise
- Femme, fille ou homme au foyer
- Retraité(e), préretraité(e), retiré(e) des affaires
- Autre inactif

**15** Avez-vous déjà travaillé dans le passé comme salarié ou à votre compte ? Oui — Quelle était votre profession principale ? / Non

**16** Faites-vous des démarches pour trouver du travail ?
- Vous n'effectuez aucune démarche
- Vous effectuez des démarches depuis :
  - moins de trois mois
  - trois mois à moins d'un an
  - un an à moins de deux ans
  - deux ans ou plus

**17** Êtes-vous disponible immédiatement pour occuper un emploi s'il se présente ? Oui / Non — Allez à la question **20**

**18** Quelle profession exercez-vous ? Soyez très précis : cultivateur de cultures vivrières, épicier détaillant, ouvrier électricien, d'entretien, maçon, secrétaire de mairie, juriste, gendarme, comptable, etc.

**19** Exercez-vous cette profession :
- de façon continue ?
- de façon intermittente ou saisonnière ?
- de façon exceptionnelle ?

**20** Dans votre emploi, êtes-vous :
- ouvrier, catégorie D ou 5 de la fonction publique ?
- secrétaire, employé de bureau, de l'hôtellerie, vendeur, agent de service, aide soignant, catégorie C ou 3 ou 4 de la fonction publique ?
- instituteur, infirmier, cat. B ou 2 de la fonction publique ?
- agent de maîtrise, maîtrise administrative ou commerciale, VRP ?
- profession libérale, ingénieur, cadre, catégorie A ou 1 de la fonction publique, ingénieur, cadre d'entreprise ?
- chef d'entreprise, artisan, commerçant, gérant d'entreprise ?
- pêcheur, aquaculteur, chasseur ?
- agriculteur exploitant, agriculteur éleveur ?
  - Si vous êtes agriculteur exploitant, êtes-vous propriétaire des terrains de l'exploitation ? Oui / Non

**21** Occupez-vous votre emploi :
- à temps complet
- à temps partiel : mi-temps et plus ? / moins que le mi-temps ?

**22** Où travaillez-vous ?
- Commune, commune associée, île
- Établissement, nom ou raison sociale
- Adresse géographique
- Activité de l'établissement qui vous emploie ou que vous dirigez

**23** Êtes-vous :
- salarié du secteur privé à durée déterminée ?
- salarié du secteur privé à durée indéterminée ?
- salarié du secteur public à durée déterminée ?
- salarié du secteur public à durée indéterminée ?
- stagiaire rémunéré ou sous contrat aidé (CPIA, ...) ?
- aide familial ?
- travailleur indépendant (artisan, commerçant, chef d'entreprise) ?
- membre d'une profession libérale ?
- agriculteur, éleveur, chasseur, pêcheur, aquaculteur ?
  - L'essentiel de la production est-il réservé à la consommation familiale ? Oui / Non

**24** Avez-vous une activité annexe ? Si tout ou partie de la production est vendue, cochez « Pour la vente » (Non / Oui / Pour la vente)
- Agriculture
- Pêche
- Artisanat
- Petit commerce de rue ou commerce à domicile
- Prestations liées à des activités touristiques
- Autres (petits boulots...)
  - Précisez

Date : ________  Signature : ________

[4] : Questionnaire de l'ISPF, feuille de logement, 2007

[5] : <u>Questionnaire réalisé à destination des pêcheurs de Moorea</u>

# Questionnaire à destination des pêcheurs de Moorea

'Ia ora na,

Je suis en stage au CRIOBE à Moorea et je réalise un modèle du comportement des pêcheurs. Mon objectif est de proposer des pistes pour s'assurer du maintien de la pêche et des poissons sur le long terme. J'ai réalisé un court questionnaire, entièrement anonyme. Si vous pouviez y répondre, cela m'aiderait beaucoup. Cela ne prend pas plus de 5 minutes.

Mauruuru roa !

Olivier Rousselle

**Pourquoi pêchez-vous ? (plusieurs réponses possibles)**
- [ ] Pour vendre
- [ ] Pour manger
- [ ] Pour la plaisir
- [ ] Autre...

**En général, à quelle distance depuis votre point de départ parcourez-vous pour vous rendre sur votre lieu de pêche (en km) ?**

0 1 2 3 4 5 6 7 8 9 10

**Quel est votre moyen de locomotion en mer pour aller pêcher ? (plusieurs réponses possibles)**
- [ ] Nage
- [ ] Pirogue / bateau sans moteur
- [ ] Bateau à moteur
- [ ] Autre...

**En général, combien de temps consacrez-vous à votre activité de pêcher en mer par journée de pêche (en heures) ?**

0 1 2 3 4 5 6 7 8 9 10

**Quels sont les jours où vous pêchez le plus ?**
- [ ] Lundi
- [ ] Mardi
- [ ] Mercredi
- [ ] Jeudi
- [ ] Vendredi
- [ ] Samedi
- [ ] Dimanche

**Vous arrive-t-il de pêcher de nuit ?**
- ( ) Oui
- ( ) Non

**Lors d'une sortie de pêche, quelle quantité de poissons pêchez-vous en général (en kilogrammes) ?**

*Réponse courte*

**Pensez-vous que le nombre de pêcheurs à Moorea :**
- ( ) Augmente
- ( ) Est constant
- ( ) Diminue

**Comment expliquez-vous l'évolution du nombre de pêcheurs à Moorea ?**

*Réponse longue*

**Quel est l'effet du tourisme sur votre pratique de pêche ?**
- ( ) Positif
- ( ) Neutre
- ( ) Négatif

**Si vous deviez mettre un quota de pêche journalier (biomasse de poisson maximale qu'il est possible de pêcher) afin d'assurer une durabilité des stocks de poissons, combien mettriez-vous (en kg) ?**

*Réponse longue*

**Merci d'avoir répondu à ce questionnaire ! Si vous souhaitez laisser un commentaire, cet espace est pour vous...**

*Réponse longue*